\documentclass[10pt,journal]{IEEEtran}
\usepackage{graphicx}
\usepackage{orcidlink}
\usepackage{float}
\usepackage{bm}
\usepackage{subfigure}
\usepackage{amsmath,amsfonts}
\usepackage{mathtools}
\usepackage{algorithm}
\usepackage{algpseudocode}
\usepackage[normalem]{ulem}
\usepackage[tableposition=top]{caption}
\usepackage{url}
\usepackage{booktabs}

\usepackage{array}
\usepackage{cite}
\usepackage{color}
\usepackage{cleveref}
\usepackage{geometry}
\usepackage{amssymb}
\usepackage{cuted}
\usepackage{lipsum} 
\usepackage{multicol}
\newgeometry{left=1.5cm,right=1.5cm,bottom=1.3cm,top=1.28cm}

\begin{document}

\title{OTFS-ISAC System with Sub-Nyquist ADC Sampling Rate}

\author{
Henglin Pu$^{\orcidlink{0009-0008-2390-5722}}$, 
Xuefeng Wang$^{\orcidlink{0000-0002-6536-4206}}$, 
Ajay Kumar$^{\orcidlink{0000-0003-1650-7415}}$, 
Lu Su$^{\orcidlink{0000-0001-7223-543X}}$, \IEEEmembership{Member, IEEE},  
Husheng Li$^{\orcidlink{0009-0009-5699-9558}}$, \IEEEmembership{Senior Member, IEEE}

\thanks{Henglin Pu, Lu Su and Husheng Li are with the Elmore Family School of Electrical and Computer Engineering, Purdue University, West Lafayette, Indiana, USA-47907 (email:
 pu36@purdue.edu, lusu@purdue.edu, husheng@purdue.edu).}%
\and
\thanks{Xuefeng Wang and Husheng Li are with the School of Aeronautics and
 Astronautics, Purdue University, West Lafayette, Indiana, USA-47907 (email:
 wang6067@purdue.edu, husheng@purdue.edu).}%
 \and
\thanks{A. Kumar is with the Department of Electrical Communication Engineering, Indian Institute of Science, Bangalore, India-560012
 (e-mail: ajaykumar124@iisc.ac.in).}%
}

\maketitle

\thispagestyle{empty}

\IEEEpeerreviewmaketitle

\begin{abstract}
Integrated sensing and communication (ISAC) has emerged as a pivotal technology for next-generation wireless communication and radar systems, enabling high-resolution sensing and high-throughput communication with shared spectrum and hardware. However, achieving a fine radar resolution often requires high-rate analog-to-digital converters (ADCs) and substantial storage, making it both expensive and impractical for many commercial applications. To address these challenges, this paper proposes an orthogonal time frequency space (OTFS)-based ISAC architecture that operates at reduced ADC sampling rates, yet preserves accurate radar estimation and supports simultaneous communication. The proposed architecture introduces pilot symbols directly in the delay-Doppler (DD) domain to leverage the transformation mapping between the DD and time-frequency (TF) domains to keep selected subcarriers active while others are inactive, allowing the radar receiver to exploit under-sampling aliasing and recover the original DD signal at much lower sampling rates. To further enhance the radar accuracy, we develop an iterative interference estimation and cancellation algorithm that mitigates data symbol interference. We propose a code-based spreading technique that distributes data across the DD domain to preserve the maximum unambiguous radar sensing range. For communication, we implement a complete transceiver pipeline optimized for reduced sampling rate system, including synchronization, channel estimation, and iterative data detection. Experimental results from a software-defined radio (SDR)-based testbed confirm that our method substantially lowers the required sampling rate without sacrificing radar sensing performance and ensures reliable communication.
\end{abstract}


\section{Introduction}

Over the past decade, integrated sensing and communication (ISAC)~\cite{ISAC1,ISAC2,ISAC3} has evolved into one of the major themes for next-generation wireless networks, particularly in the context of 6G systems. ISAC unifies sensing and communication within a single system infrastructure, leveraging shared spectral and hardware resources. This integration simplifies system deployments and enables transformative applications such as environment monitoring~\cite{environment1,environment2}, autonomous driving~\cite{autonomous_driving1,autonomous_driving2}, human-computer interaction~\cite{HCI1,HCI2}, and smart city initiatives~\cite{smartcity1,smartcity2}.

Thanks to significant advancements in hardware technology, substantial progress in ISAC has been achieved in the development of multiple-input-multiple-output (MIMO) technology~\cite{MIMO} and the expansion of frequency bandwidths~\cite{terahertz}. In modern 5G communication systems, MIMO technology is fully leveraged to exploit spatial resources, resulting in notable improvements in spectral and energy efficiency. Similarly, MIMO radar benefits from these advancements by enabling finer angle estimation and enhanced sensing capabilities~\cite{MIMO_radar}. Moreover, hardware improvements have facilitated the extension of the radio frequency spectrum from the frequency bands used in 4G systems to millimeter-wave (mmWave) and terahertz (THz) bands. These higher frequency bands offer broader bandwidths, significantly enhancing communication throughput and radar sensing resolution. However, these advances introduce a critical challenge for ISAC waveform design. As bandwidth grows, the Nyquist-Shannon theorem~\cite{Nyquist_shannon} requires ever-higher sampling rates to avoid signal distortion, thereby placing substantial demands on analog-to-digital converters (ADCs). In radar systems that use predefined waveforms, such as frequency-modulated continuous wave (FMCW), stretch processing can be applied to downconvert the high-bandwidth echo to an intermediate frequency (IF) signal, thereby reducing the required sampling rate before digitization. In contrast, communication signals contain unknown data and need full-band Nyquist sampling for effective decoding, making the integration of radar and communication waveforms more complex. For instance, according to the Nyquist-Shannon theorem, achieving GHz-level bandwidth transmission without distortion requires
 an ADC capable of sampling at gigasamples per second (GSPS). The commonly used 12-bit AD9625 ADC~\cite{ADC}, with a sampling
 rate of 2.6 GSPS, is priced at approximately $\$1,500$ per unit. This makes it prohibitively expensive for systems that require multiple ADCs in multi-channel configurations. Operating at this rate, a single ADC generates a data flow of 31.2 Gbps, necessitating advanced storage and data transmission solutions. This, in turn, escalates the overall cost of commercial ISAC systems and restricts their affordability and broader adoption. 
\setlength{\parskip}{0pt} 



Currently, FMCW radar remains the predominant waveform in radar applications. It achieves a high resolution using a limited ADC through the stretch processing. However, the FMCW waveform offers a limited flexibility in encoding the transmission data~\cite{FMCW_comms1,FMCW_comms2}, confining its applications to scenarios with low transmission rates such as LoRa~\cite{LoRa1,Lora2} and chirp spread spectrum (CSS)~\cite{CSS} schemes. As the need for high throughput ISAC applications grows, there has been a shift towards orthogonal frequency division multiplexing (OFDM)~\cite{OFDM1,OFDM2}. OFDM is a well-established and proficient modulation scheme extensively adopted in recent wireless communication standards, including 4G LTE and 5G NR. Numerous radar systems based on OFDM have been proposed, highlighting its potential as a waveform for ISAC applications. Nonetheless, OFDM faces substantial challenges in high mobility environments, primarily due to Doppler shifts. To address this issue, OTFS~\cite{OTFS1,OTFS2,OTFS3} modulation has been emerged as a viable solution, noted for its robustness in maintaining consistent performances under doubly-dispersive channels.

In this paper, we present an OTFS-based ISAC system that operates at a reduced ADC sampling rate without sacrificing radar sensing resolution and range. We first strategically place pilot symbols in the delay-Doppler (DD) domain according to a carefully derived transformation mapping between DD and TF domains. By doing so, we ensure that only certain subcarriers in the time-frequency (TF) domain carry active pilot symbols, while all others remain inactive. This selective activation facilitates a structured aliasing effect when undersampling is applied, allowing signals from multiple sub-bands to be effectively merged into a narrower bandwidth. To accurately retrieve the target's radar information from the downsampled signals, we introduce a two-phase iterative cancellation and detection algorithm. For data modulation and demodulation within the constraints of Nyquist-Shannon sampling theorem, effective data is allocated at the transmitter to maximize capacity according to the reduced sampling rate at the receiver. A code-based spreading technique is then applied to distribute the data across the entire DD domain, ensuring the maximum radar estimation range. Furthermore, we design a complete synchronization, modulation and demodulation pipeline that accommodates the reduced sampling rate at the receiver, and we adopt an iterative channel estimation and data detection algorithm to reliably decode the transmitted information. This ISAC system achieves a precise estimation with a sub-Nyquist sampling rate, thereby reducing hardware complexity and data storage requirements, making it highly suitable for low-budget commercial applications. We build an OTFS-based ISAC prototype using software-defined radios (SDRs). Through simulations and extensive experiments, we demonstrate the effectiveness of our system with reduced sampling rate, which delivers accurate radar estimation without compromising resolution or sensing range, while also ensuring reliable communication. In summary, this paper the following contributions:
\begin{itemize}
    \item As per the authors' best knowledge, this work is the first to design and implement an OTFS-based ISAC system that operates with a reduced ADC sampling rate, addressing the cost and complexity concerns while maintaining a high radar performance;
    \item We propose an OTFS-based radar estimation algorithm that achieves precise sensing without sacrificing the resolution or the maximum unambiguous sensing range, even under reduced sampling rate constraints;
    \item We develop an integrated communication pipeline, including modulation, demodulation, synchronization, channel estimation, and iterative data detection, ensuring a reliable data transmission in the reduced sampling rate framework.
    \item We have implemented the proposed system on our ISAC testbed using SDRs and conducted extensive experimental studies. The results validate the effectiveness of our approach, demonstrating accurate radar estimation and robust communication performance under reduced sampling rates, without compromising the sensing resolution or communication performance.
\end{itemize}
The remainder of
this paper is organized as follows. The researches related to this
paper are introduced and compared in Section \ref{sec:related}. The general
system model is introduced in Section \ref{sec:system}. Then,
the radar estimation algorithm and communication pipeline are elaborated in
Section \ref{sec:radar} and \ref{sec:comms} respectively. The properties of the reduced sampling rate system compared with the case of full sampling rate are discussed in Section \ref{sec:properties}. The corresponding numerical
and experimental results are provided in Section \ref{sec:results}. Finally the
conclusions are drawn in Section \ref{sec:conclusion}.

\section{Related Works}\label{sec:related}

Several strategies have been developed to reduce the ADC sampling rate in OFDM radar systems. The stepped-carrier (SC) OFDM radar~\cite{SC-OFDM} divides a wide-bandwidth OFDM signal into narrower sub-bands, transmitting them sequentially, which affects the maximum unambiguous velocity and requires a fast-settling-time phase-lock loop (PLL) and phase-offset calibration between sub-bands. Sparse OFDM radar~\cite{compressive} utilizes narrow-band signals at randomly chosen frequencies, reconstructed via compressed sensing to match the full OFDM resolution, though it increases the hardware complexity and computational demand. Frequency-comb (FC) OFDM radar~\cite{FC1,FC2} expands the RF bandwidth by multiplying the baseband signal with a frequency comb, at the cost of complicating the hardware due to its need for precise calibration to avoid peak-to-sidelobe ratio degradation. Subcarrier-aliasing (SA) OFDM radar~\cite{SA} designs a proper active subcarrier interval in the OFDM signal allows active subcarriers in each sub-band not to be overlapped in the undersampled
signal. Although this involves simple hardware and radar
signal processing compared to other approaches, its system
performances in terms of the maximum unambiguous range
and the range processing gain are severely degraded due to
the reduced number of active subcarriers. Recently, a sub-Nyquist sampling (SNS)
OFDM radar system with reduced sampling rate is proposed in~\cite{SNS}, which ensures a high range resolution and the maximum detectable range without ambiguities. The proposed method uses sub-Nyquist sampling, with demodulation unfolding the signal and introducing symbol-mismatch noise (SMN). The SMN is then canceled, thus restoring the dynamic range of the unfolded signal. However, the effectiveness of this approach relies heavily on the assumption that the entire bandwidth is fully occupied by radar symbols. This assumption limits the feasibility of the approach in ISAC applications.

In OTFS modulation, an OTFS-FMCW waveform design is proposed in~\cite{OTFS-FMCW}. This design capitalizes on the simultaneous locality property of the FMCW in both the time-frequency and delay-Doppler domains to superimpose OTFS and FMCW signals in an orthogonal manner. However, the authors obscure a critical concept pertinent to multicarrier modulations such as OTFS and OFDM, wherein each specific tone remains static during its transmission interval and alters only between successive intervals. This attribute aligns the approach more closely with stepped-frequency radar rather than with continuous-wave frequency modulation. Consequently, it cannot straightforwardly harness the intrinsic properties of FMCW radar.

\section{System Overview}\label{sec:system}

\begin{figure}[!t]
 \centering
 \includegraphics[width=3.5in]{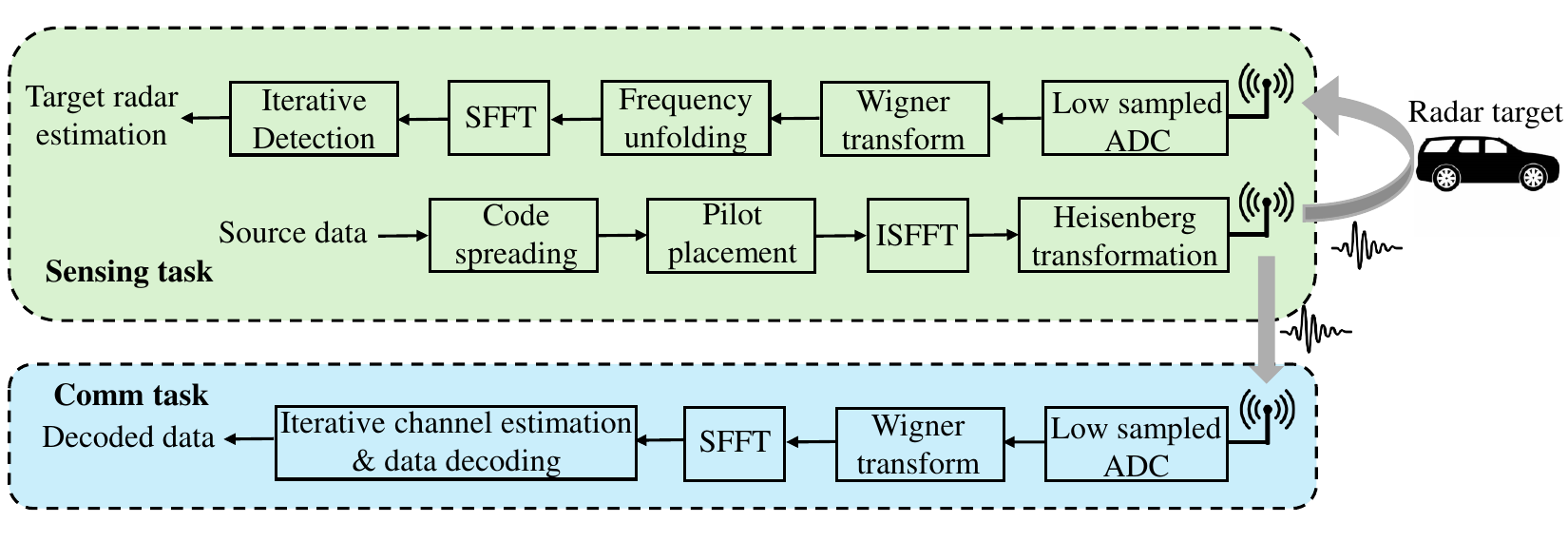}\\
 \caption{Processing pipeline of OTFS-ISAC system with reduced ADC sampling rate.}
 \label{fig:Illustration}
\end{figure}

In this section, we provide the overview of the OTFS-ISAC system and the design of the proposed pilot scheme, taking into account the effects of reduced sampling rate aliasing. As depicted in Fig.~\ref{fig:Illustration}, the OTFS-ISAC system can be principally divided into two functional areas in terms of application tasks, namely the radar sensing part and the communication part. In the system, the transmitter operates at full sampling rate, while the receivers, both on the radar and communication ends, operate at reduced sampling rates. This configuration ensures sufficient physical bandwidth on the transmitter side for the accurate radar parameter estimation, while the reduced sampling rate at the receivers helps mitigate the stringent demands on the ADC and data storage capacities.

\subsection{Input-Output Relation of OTFS}
In this section, the signal model for OTFS modulation is delineated. The transmission data $\mathbf{X}^{DD} \in \mathbb{C}^{M_{\tau}\times N_{\nu}}$ is embedded in the DD domain with an $M_{\tau}\times N_{\nu}$ grid. Each grid cell spans $T$ seconds in the time dimension and $\Delta_f$ Hertz in the Doppler shift, uniquely corresponding to a specific delay and Doppler pair. $M_{\tau}$ and $N_{\nu}$ denote the number of delay and Doppler bins respectively. Then by applying an Inverse Symplectic Finite Fourier Transform (ISFFT), the DD domain signals are transformed into the TF domain:
\begin{equation}\label{eq:ISFFT}
\begin{array}{l}
\mathbf{X}^{TF} = \mathbf{F}_{M_{\tau}}\mathbf{X}^{DD} \mathbf{F}^H_{N_{\nu}}, 
\end{array}
\end{equation}
where $\mathbf{X}^{TF} \in \mathbb{C}^{M_{\tau}\times N_{\nu}}$ is the time-frequency domain representation and $\mathbf{F}_N$ is the unitary discrete Fourier transform (DFT) matrix of size $N\times N$. Subsequently, the Heisenberg transform is employed and a transmit pulse-shaping filter $g_{tx}(t)$ supported in $[0,T]$ is utilized to generate the transmission signal $s(t)$ in time domain:
\begin{equation}
\begin{array}{l}
s(t) = \frac{1}{\sqrt{M_{\tau}}} \sum\limits^{N_{\nu}-1}\limits_{n=0} \sum\limits^{M_{\tau}-1}\limits_{m=0} \mathbf{X}^{TF}[m,n] g_{tx}(t-nT)\\
\qquad \qquad \qquad \qquad \qquad \qquad \qquad \quad e^{j2\pi m\Delta_f(t-nT)}.
\end{array}
\end{equation}

On the communication receiver side, the received signal $r(t)$ is multiple delayed copies of transmitted ones: 
\begin{equation}
\begin{array}{l}
r(t) = \int_{\nu} \int_{\tau} h(\tau, \nu) s(t-\tau) e^{j2\pi \nu(t-\tau)}d\tau d\nu + w(t)\\
\qquad = \sum\limits^{P}\limits_{i=1} h_i e^{j2\pi v_i t} s(t-\tau_i) + w(t),
\end{array}
\end{equation}
where $w(t)$ represents the additive white Gaussian noise, $P$ denotes the number of paths, $h(\tau, \nu)$ represents the channel impulse response:
 \begin{equation}\label{eq:channelH}
\begin{array}{l}
h(\tau ,v) = \sum\limits^{P}\limits_{i=1} h_i \delta(\tau - \tau_i) \delta(\nu - \nu_i),
\end{array}
\end{equation}
where $h_i$ is the complex gain associated with the $i$-th path, while $\delta (\cdot)$ is the Dirac delta function. The expression for path delay $\tau_i$ and Doppler shift $\nu_i$ of the $i$-th path assumes integer multiples of the respective delay and Doppler resolutions, articulated as $\tau_i = \frac{l_i}{M_{\tau}\Delta f}$ and $\nu_i =\frac{k_i}{N_{\nu}T}$, respectively. Here, $l_i$ and $k_i$ are integers denoting the delay and Doppler taps of the $i$-th path, respectively. 
This time domain signal $r(t)$ is sampled and transformed to the TF domain signal $\mathbf{Y}^{TF} \in \mathbb{C}^{M\times N}$ by applying the Wigner transform: 
\begin{equation}
\begin{array}{l}
\mathbf{Y}^{TF}[m,n] = \int g^*_{rx}(t-nT)r(t)e^{-j2\pi m\Delta_f (t-nT)}dt, 
\end{array}
\end{equation}
where $g_{rx}(t)$ is the receive pulse-shaping filter. Next, the Symplectic Finite Fourier Transform (SFFT) is applied to transform $\mathbf{Y}^{TF}$ back to the DD domain:
\begin{equation}\label{eq:SFFT}
\begin{array}{l}
\mathbf{Y}^{DD} = \mathbf{F}^{H}_{M_{\tau}
} \mathbf{Y}^{TF} \mathbf{F}_{N_{\nu}}.
\end{array}
\end{equation}

Combining Eqs.~(\ref{eq:ISFFT})-(\ref{eq:SFFT}), the input-output relationship of OTFS in DD domain can be written as
\begin{equation}\label{eq:IO}
\begin{aligned}
\mathbf{y} &= H\mathbf{x} + \mathbf{w} \\
&= \sum\limits^{P}\limits_{i=1} h_i \Gamma_i(\tau_i, v_i) \mathbf{x} + \mathbf{w}, 
\end{aligned}
\end{equation}
where $\mathbf{x}^{DD}=vec(\mathbf{X}^{DD})$ and $\mathbf{y}^{DD}=vec(\mathbf{Y}^{DD})$. The noise vector $\mathbf{w}\sim \mathcal{N}_c(\mathbf{0},\sigma_{w}^2\mathbf{I})$, where $\mathbf{I}$ denotes the identity matrix. $\Gamma_i$ denotes the DD domain channel matrix of each reflected path:
\begin{equation}\label{eq:Gamma}
\begin{array}{l}
\Gamma_i(\tau_i, v_i) = (\mathbf{F}_{N_{\nu}} \otimes \mathbf{G}_{rx}) \Pi^{l_i}_{M_{\tau}N_{\nu}} \Delta_{M_{\tau}N_{\nu}}^{k_i} (\mathbf{F}^H_{N_{\nu}} \otimes \mathbf{G}_{tx}),
\end{array}
\end{equation}
where $\Delta$=diag$\{ e^{j 2\pi \frac{0}{M_{\tau}N_{\nu}}}, e^{j 2\pi \frac{1}{M_{\tau}N_{\nu}}}, \cdots , e^{j 2\pi \frac{M_{\tau}N_{\nu}-1}{M_{\tau}N_{\nu}}}\}$. The matrix $\Pi_{M_{\tau}N_{\nu}}$ denotes the forward cyclic-shift (permutation) matrix. $\mathbf{G}_{tx}$ and $\mathbf{G}_{rx}$ are the $M_{\tau}\times M_{\tau}$ diagonal matrices whose entries are the coefficients of the corresponding transmit pulse and receive pulse (in this paper we make the assumption that both the transmit and receive pulses are rectangular, thus that $\mathbf{G}_{tx}$ and $\mathbf{G}_{rx}$ are identity matrices, denoted $\mathbf{I}_{M_{\tau}}$).

\begin{figure}[!t]
 \centering
 \includegraphics[width=2.6in]{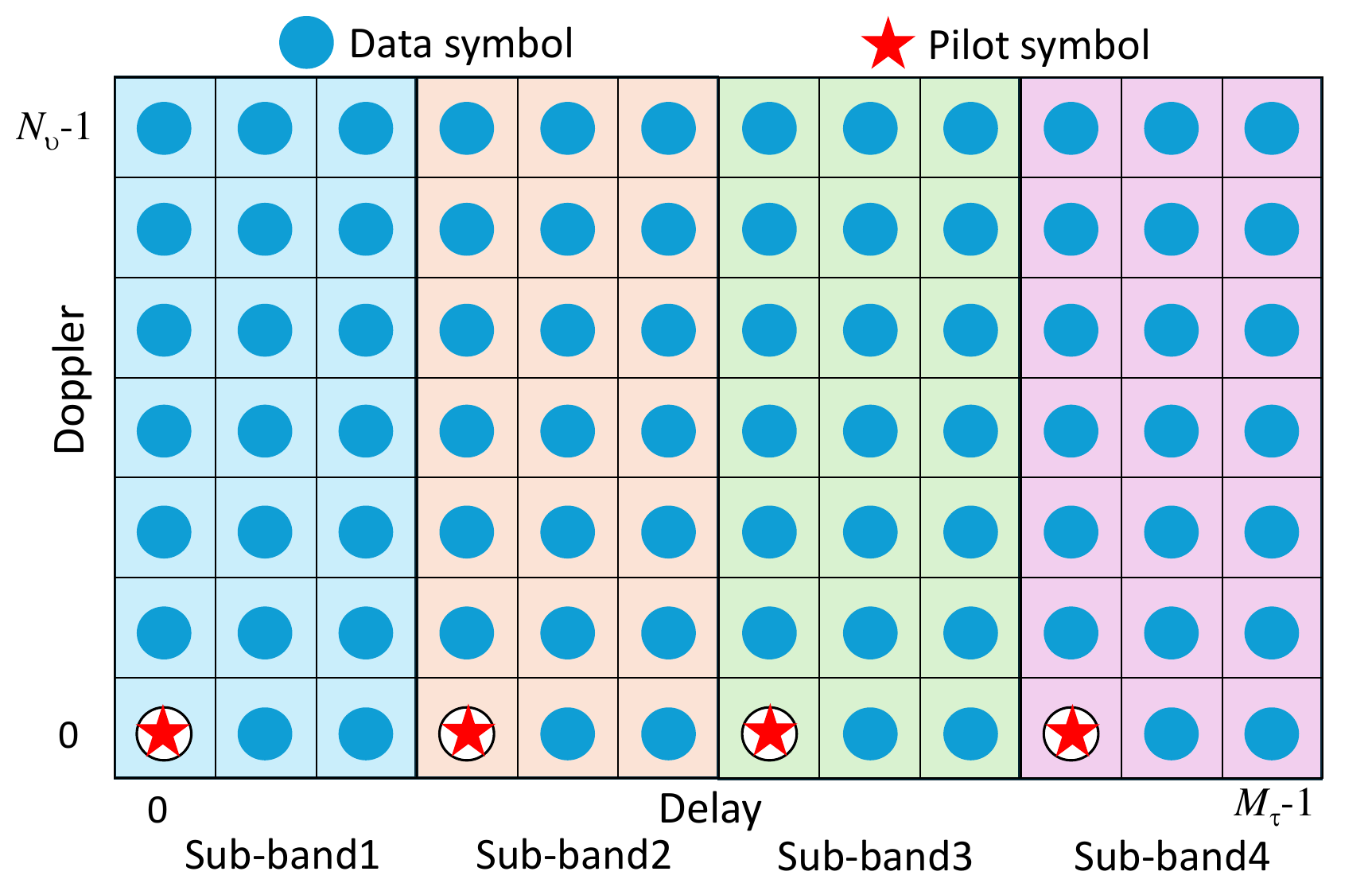}\\
 \caption{Pilot placement in the DD domain.}
 \label{fig:pilot_placement}
\end{figure}

\begin{figure*}[!h]
    \centering
    \subfigure[]
    {
        \label{fig:response:aa}
        \includegraphics[width=1.55in]{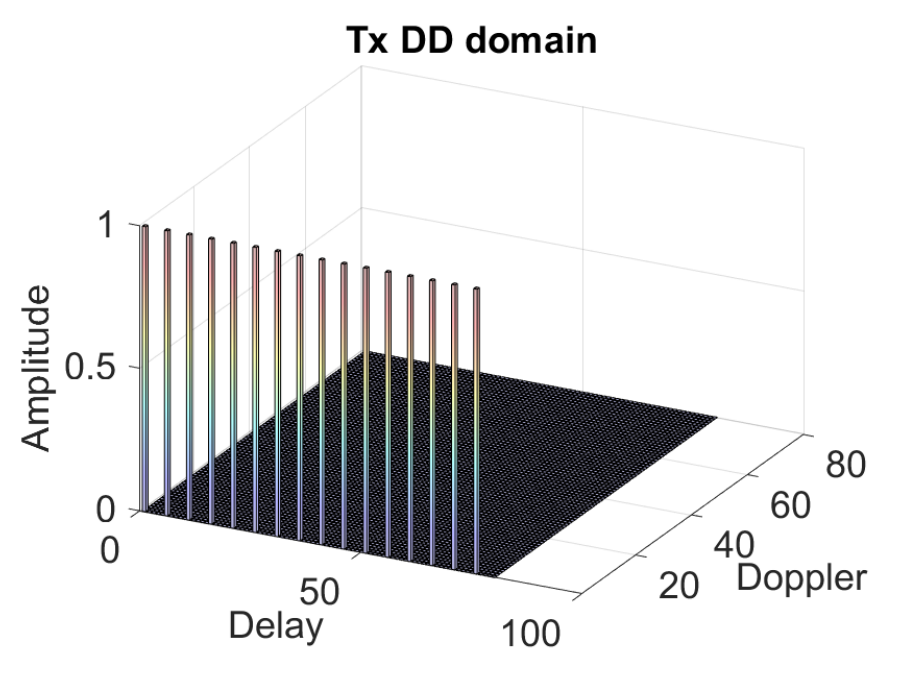}
    }
    \hspace{0.0001\linewidth}
    \subfigure[]
    {
       \label{fig:response2:a}
        \includegraphics[width=1.55in]{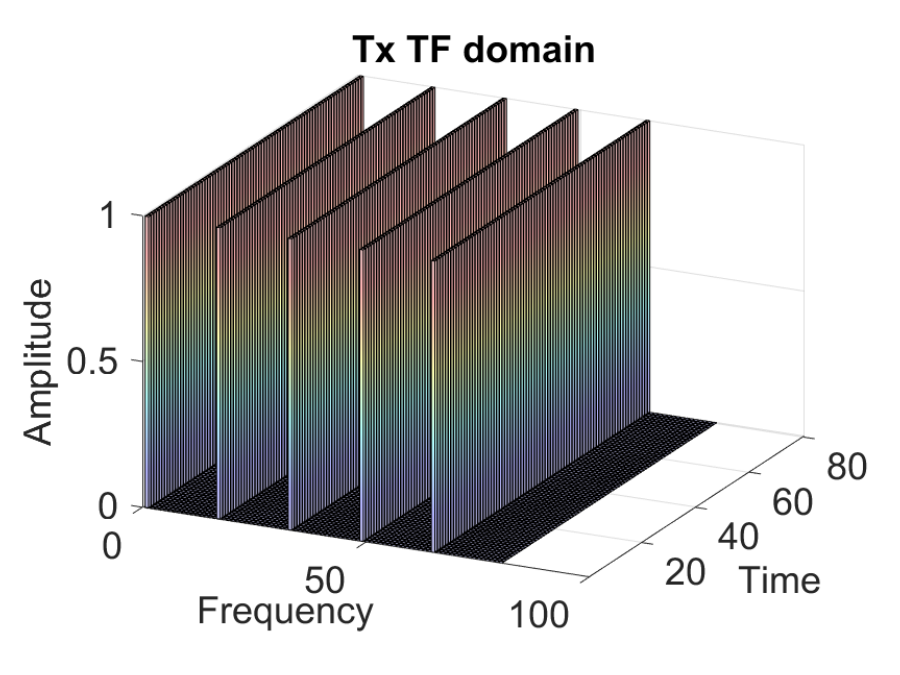}
    }
    \hspace{0.0001\linewidth}
    \subfigure[]
    {
        \label{fig:response:b}
        \includegraphics[width=1.55in]{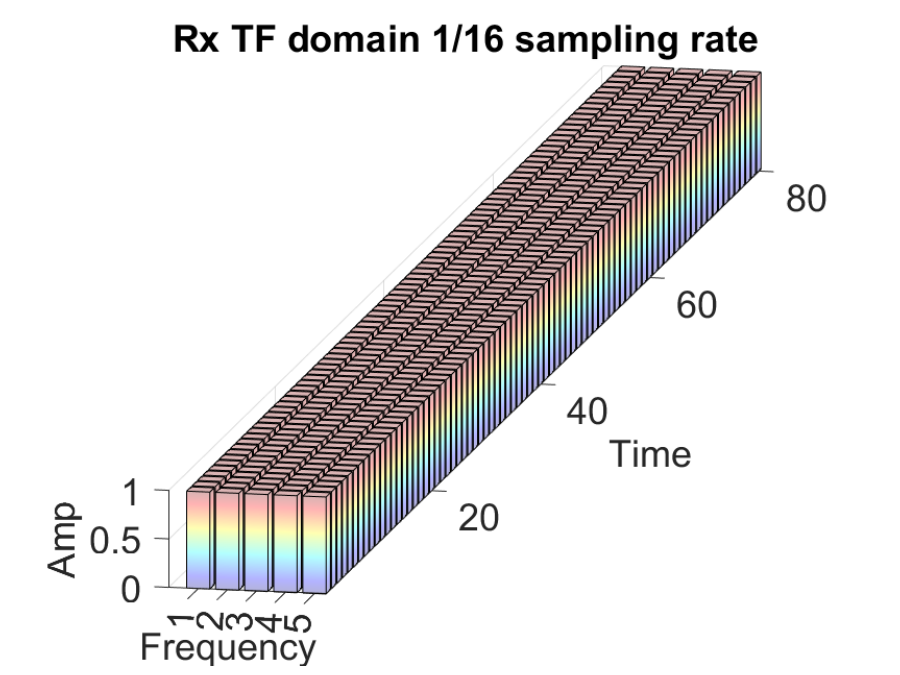}
    }
    \hspace{0.0001\linewidth}
    \subfigure[]
    {
        \label{fig:response:c}
        \includegraphics[width=1.55in]{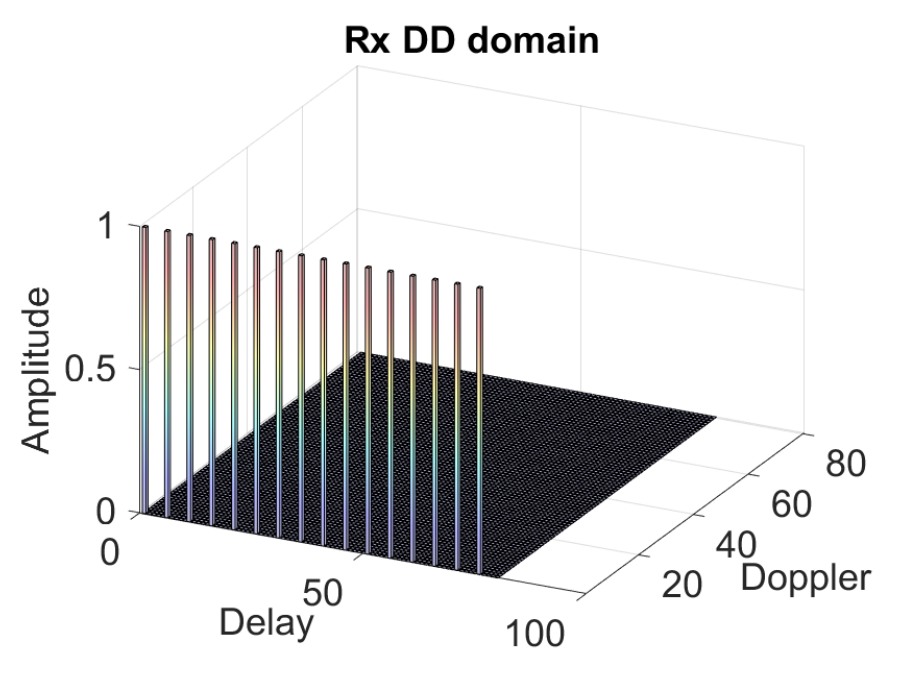}
    }
    
\caption{Representations of pilot signal in different domains: (a) transmitted pilot signal in DD domain, (b) transmitted pilot signal in TF domain, (c) downsampled captured pilot signal in TF domain, (d) Recovered pilot signal in DD domain.}
\label{fig:overall performance of waveform recovery}
\end{figure*}

\subsection{OTFS-based Radar}

In the monostatic OTFS-based radar system, the transmit symbols $\mathbf{X}^{DD}$ and the received symbols $\mathbf{Y}^{DD}$ are known.
However, the round-trip delay–Doppler radar channel, denoted as $h_r(\tau, \nu)$, remains unknown. The purpose of radar processing is to
estimate $(h_r,\tau,\nu)$, which provides us the target information, such as the corresponding ranges and velocities. The commonly used model for OTFS radar parameter estimation based on maximum likelihood estimation (MLE)~\cite{OTFS_radar1,OTFS_radar2} is described as follows.

For the $i^{\text{th}}$ target, let $h_{r,i}$ denote the corresponding round-trip channel coefficient. When estimating the parameters of the $i^{\text{th}}$ target, the reflections from other targets should be treated as interference. Specifically, $h_{r,i} \Gamma_i \mathbf{x}$ and $\sum_{j \neq i} h_{r,i} \Gamma_j \mathbf{x}$ represent the signal and interference, respectively. To address the mutual interference, an interference cancellation mechanism is required. When estimating the parameters of the $i^{\text{th}}$ target, the interference from the $(i-1)$ previously estimated targets must be removed. Considering the first target with parameters $(h_{r,1}, \tau_1, \nu_1)$, The MLE for this target is given by
\begin{equation}
(\hat{h}_{r,1}, \hat{\tau}_1, \hat{\nu}_1) = \arg \min_{(h_{r,1}, \tau_1, \nu_1)} \| h_{r,1} \Gamma_1 \mathbf{x} - \mathbf{y} \|^2.
\end{equation}
Next, the above minimization problem can be reformulated as a maximization problem as
\begin{equation}
(\hat{\tau}_1, \hat{\nu}_1) = \arg \max_{(\tau_1, \nu_1)} \left| (\Gamma_1 \mathbf{x})^H \mathbf{y} \right|^2,
\end{equation}
and the channel coefficient $h_{r,1}$ is calculated by $\hat{h}_{r,1} = \frac{\mathbf{x}^H \Gamma_1^H (\hat{\tau}_1, \hat{\nu}_1) \Gamma_1 (\hat{\tau}_1, \hat{\nu}_1) \mathbf{x}}{(\Gamma_1 (\hat{\tau}_1, \hat{\nu}_1) \mathbf{x})^H \mathbf{y}}$. For the remaining $(P-1)$ targets, such as the $i^{\text{th}}$ target, where $i = 2, 3, \dots, P$, interference cancellation is performed by subtracting the interference term $\sum_{j=1}^{i-1} \hat{h}_{r,j} \Gamma_j (\hat{\tau}_j, \hat{\nu}_j) \mathbf{x}$ from the received vector $\mathbf{y}$. Accordingly, the estimator for the $i^{\text{th}}$ target is expressed as
\begin{equation}\label{eq:MLE_tau_mu}
(\hat{\tau}_i, \hat{\nu}_i) = \arg \max_{(\tau_i, \nu_i)} \left| (\Gamma_i \mathbf{x})^H \left( \mathbf{y} - \sum_{j=1}^{i-1} \hat{h}_{r,j} \Gamma_j (\hat{\tau}_j, \hat{\nu}_j) \mathbf{x} \right) \right|^2,
\end{equation}
and the estimated channel coefficient $h_{r,i}$ is derived as:
\begin{equation}\label{eq:MLE_alpha}
\hat{h}_{r,i} = \frac{\mathbf{x}^H \Gamma_i^H (\hat{\tau}_i, \hat{\nu}_i) \Gamma_i (\hat{\tau}_i, \hat{\nu}_i) \mathbf{x}}{(\Gamma_1 (\hat{\tau}_1, \hat{\nu}_1) \mathbf{x})^H \left( \mathbf{y} - \sum_{j=1}^{i-1} \hat{h}_{r,j} \Gamma_j (\hat{\tau}_j, \hat{\nu}_j) \mathbf{x} \right)}.
\end{equation}

\begin{figure}[!t]
 \centering
 \includegraphics[width=3.5in]{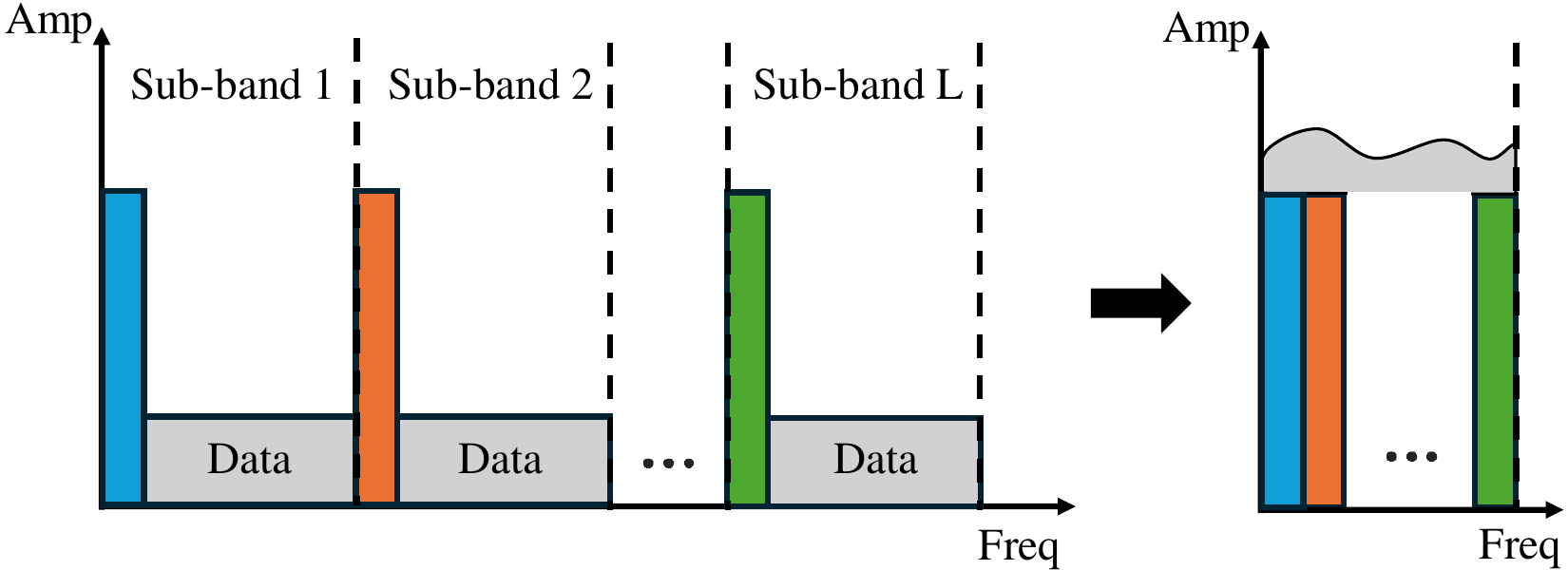}\\
 \caption{Illustration of reduced sampling rate aliasing.}
 \label{fig:aliasing}
\end{figure}

\subsection{Pilot Scheme for Reduced Sampling Rate System}

Inspired by the subcarrier-aliasing technique introduced in OFDM radar systems~\cite{SA}, we propose to generate a DD domain signal by strategically placing pilot signals in designated grids within the DD domain. This configuration ensures that these pilot signals correspond to active symbols on every $\mu$th subcarrier in the TF domain. Utilizing the aliasing effect, this method efficiently folds the full bandwidth signal into a sub-band without loss of pilot information. Consider the full sampling rate $f_s$ at the transmitter, and the downsample rate at the receiver side is $\kappa$. The captured signal is folded into $\frac{1}{\kappa}$ of its original bandwidth, while the time axis remains unchanged. Unlike OFDM, which intuitively modulates data directly in the TF domain, allowing straightforward design and control of aliasing, OTFS is tailored to combat doubly-dispersive channels by spreading each DD domain symbol across the entire TF domain. Consequently, designing a pilot scheme for OTFS requires strategic planning and poses non-trivial challenges for subsequent radar estimation and communication processing.

From Eq.~(\ref{eq:ISFFT}), the time-domain representation of $\mathbf{X}^{TF}$ is derived from the inverse DFT applied to the DD domain signal $\mathbf{X}^{DD}$ along the Doppler axis. Similarly, the frequency-domain representation of $\mathbf{X}^{TF}$ is obtained by performing the DFT on $\mathbf{X}^{DD}$ along the delay axis. By carefully placing pilot impulses along delay axis in the DD domain and controlling their periodic spacing, we induce a structured aliasing in the TF domain that can be later exploited to recover range-Doppler estimates with a reduced sampling rate.

For reference, consider the DFT of a finite-length sequence $x[n]$ defined for $0 \le n \le N-1$:
\begin{equation}\label{eq:DFT_formula}
X[k] = \sum_{n=0}^{N-1} x[n] e^{-i\frac{2\pi kn}{N}}.
\end{equation}

When input $x[n]$ is an impulse $\delta[n]$ then Eq.~(\ref{eq:DFT_formula}) is reduces to
\begin{equation}
X[k] = \sum_{n=0}^{N-1} \delta[n] e^{-i 2 \pi k n / N} = 1.
\end{equation}
Thus, the DFT of a single impulse is a constant spectrum of 1 across all $k$.

Now consider a periodic impulse sequence:
\begin{equation}
x[n] = \sum_{k=-\infty}^{\infty} \delta[n - kT_p],
\end{equation}
where \( \delta[n] \) is the Kronecker delta function, and \( T_p \) is the period of the impulse sequence. The DFT of \( x[n] \) is given by
\begin{equation}
    \begin{aligned}
        X[k] &= \sum_{n=0}^{N-1} \left(\sum_{m=-\infty}^{\infty} \delta[n - mT_p]\right) e^{-i 2\pi \frac{k}{N} n},
    \end{aligned}
\end{equation}

Since \( \delta[n - mT_p] = 1 \) for \( n = mT_p \) (and zero otherwise), and considering \( n < N \), the equation is simplified to
\begin{equation}
    \begin{aligned}
X[k] = \frac{1 - \omega^{\lfloor \frac{N-1}{T_p} \rfloor + 1}}{1 - \omega},
    \end{aligned}
\end{equation}
where  \( \omega = e^{-i 2\pi \frac{k}{N} T_p} \). When we take the DFT of $x[n]$ over an interval of $N$ samples, the impulses at multiples of $T_p$ contribute to a repetitive pattern in frequency. This repetition results in a frequency spectrum that repeats every 
$\frac{N}{T_p}$ samples. 

Based on the above derivation, our system ensures that the TF domain signals remain active only every \(\mu\)-th subcarrier by strategically placing pilot impulses in the DD domain. Specifically, we position these pilots along the delay axis at intervals of \(\frac{N_c}{\mu}\), where \(N_c\) denotes the total number of subcarriers, which is equal to the number of delay bins \((M_{\tau})\). As illustrated in Fig.~\ref{fig:pilot_placement} and \ref{fig:overall performance of waveform recovery}(a), this placement partitions the DD domain into \(\mu\) sub-bands along the delay axis, each with dimensions \(\frac{M_{\tau}}{\mu} \times N_{\nu}\). To further enhance the spectral efficiency, a single impulse pilot is allocated along the Doppler axis, thus reserving the majority of Doppler resources for data transmission. Fig.~\ref{fig:pilot_placement}(b) and (c) depict the resulting TF-domain signals observed at both the transmitter and receiver.

Unlike a pure radar task, where all symbols are dedicated to radar parameter estimation, the ISAC system allocates the majority of resources to data embedding, with only the pilot symbols reserved for estimation, as shown in Fig.~\ref{fig:pilot_placement}. In each sub-band, there is only a single impulse for the pilot, while the remaining resources are used for data transmission. After the aliasing effect incurred by the reduced sampling rate, the signals from all sub-bands are folded into a single band as shown in Fig.~\ref{fig:aliasing}. To ensure that the folded pilots remain free from mutual interference, a careful selection of system parameters is needed.

\subsection{Parameter Selection}

Consider \( x(t) \) be a complex-valued signal with a bandwidth \( B \). The Fourier transform of the sampled signal \( x_s(t) \) with a sampling rate of $f_s'$ results in a replication of the original spectrum \( X(f) \) at intervals of \( f_s' \), namely

\begin{equation}
    X_s(f) = \sum_{k=-\infty}^{\infty} X(f - k f_s'),
\end{equation}
where \( k \) is an integer representing the index of the spectral replication. The replicated spectra are shifted by multiples of \( f_s' \) in both the positive and negative frequency ranges.

If the sampling rate \( f_s' \) is less than the bandwidth \( B \), aliasing occurs. In this case, the spectral replicas overlap, and frequency components beyond the Nyquist frequency fold back into the baseband. For a complex signal containing a frequency component \( f_0 \), the aliased frequency \( f_{\text{alias}} \) can be determined by

\begin{equation}\label{eq:aliasing}
    \begin{aligned}
        f_{\text{alias}} = f_0 - k f_s' \quad \text{for positive frequencies} \\
        f_{\text{alias}} = f_0 + k f_s' \quad \text{for negative frequencies}
    \end{aligned}
\end{equation}

As the signal is complex, the positive and negative frequencies are treated independently. In our system, both the reduced sampling rate $\kappa$ and the active subcarrier frequencies, which are determined by the number of subcarriers ($M_{\tau}$) and active subcarrier spacing $\mu$, must be designed carefully. The first constraint requires that the number of subcarriers ($N_c$), which is also equal to the number of delay bins ($M_{\tau}$), be an integer multiple of $\kappa$. Otherwise, the active subcarriers would be placed “in between” the frequency bins, causing severe windowing effects due to
loss of orthogonality. The second constraint relates \(\kappa\) and \(\mu\). After the aliasing operation in Eq.~(\ref{eq:aliasing}), all active carriers should fold into a single sub-band without interference. Since the size of each sub-band is \(\tfrac{M_{\tau}}{\mu}\), and the number of subcarriers after downsampling by \(\kappa\) is \(\tfrac{M_{\tau}}{\kappa}\), these quantities must match. Equating \(\frac{M_{\tau}}{\mu}\) and \(\frac{M_{\tau}}{\kappa}\) yields \(\mu = \kappa\).

To determine suitable parameters of $M_{\tau}$ and $\kappa$, we propose a parameter search algorithm, outlined in Algorithm~\ref{alg:parameter_selection}. Specifically, the goal is to identify the optimal downsampling rate, $\kappa$, at the receivers and the number of delay bins in the DD domain. The algorithm iterates over potential values for these parameters. For each iteration, a specific pilot placement in the DD domain is constructed based on the current parameter values. The DD signals are then transformed into the TF domain, where the aliasing effect is simulated, and the resulting aliased signal in the frequency domain is analyzed. If the normalized aliased signal in the frequency domain consists of only ones, it indicates that every frequency bin is occupied by only one active subcarrier. This ensures that the chosen parameters eliminate the inter-subband interference and prevent information loss. Consequently, the appropriate parameters are identified to maintain the signal integrity. In most cases, the reduced sampling rate, $\kappa$, at the receiver is predetermined and fixed by the hardware constraints of the system. This significantly reduces the complexity of the parameter search process. 

\begin{algorithm}[t]
\caption{Parameter selection}\label{alg:parameter_selection}
\begin{algorithmic}[1]
\Require $M_{min}$, $M_{max}$, $\kappa_{\max}$
\For{$\kappa = 2, 3, \ldots, \kappa_{\max}$}
\For{$M_{\tau} = M_{min}, M_{min}+1, \ldots,M_{max}-1, M_{max}$}
\If{$\frac{M_{\tau}}{\kappa}$ $\in$ $\mathbb{Z}$}
\State $\mu = \kappa$, $f_s' = \frac{f_s}{\kappa}$;
\State Construct pilot placement scheme $\mathbf{X}_p^{DD}$ based on $M_{\tau}$ and $\mu$;
\State Construct $\mathbf{Y}_p^{TF}$ based on Eq.~(\ref{eq:SFFT});
\State Construct $\mathbf{B}_a$ based on $f_s',\mu$,$M_{\tau}$;
    \State Derive $\tilde{\mathbf{Y}}_p^{TF} = \mathbf{B}_{a}\mathbf{Y}_p^{TF} $;
\If{$\tilde{\mathbf{Y}}_p^{TF}$ contains $\frac{M_{\tau}}{\kappa}$ ones}
        \State \Return $M_{\tau}$, $\kappa$;
    \EndIf
    \EndIf  
\EndFor
\EndFor
\end{algorithmic}
\end{algorithm}

\section{Radar Parameter Estimation with Reduced Sampling Rate}\label{sec:radar}

\begin{algorithm}[!b]
\caption{The iterative radar detection algorithm}\label{alg:cap}
\begin{algorithmic}[1]
\State \textbf{Input:} Convergence error $\epsilon$, transmitted data signal $\mathbf{x}^{DD}_d$, transmitted pilot signal $\mathbf{x}^{DD}_p$, unfolded received signal $\Tilde{\mathbf{y}}^{DD}$;

\State \textbf{Output:} Target range estimation $\hat{\tau}$, target velocity estimation $\hat{\nu}$;
\State Perform the MLE estimation in Eqs.~(\ref{eq:MLE_tau_mu})(\ref{eq:MLE_alpha}) to $\mathbf{y}_{\text{uf}}^{DD}$ based on $\mathbf{x}^{DD}_p$ to obtain a coarse estimation $(\hat{h}_r,\hat{\tau},\hat{\nu})$;
\State Iterate all possible wrap around range estimations, find the absolute estimation, update the estimation $(\hat{h}_r,\hat{\tau},\hat{\nu})$;
\While{not terminate}
    \State Generate ideal received data signal $\mathbf{y}_d^{sim}$ based on $\mathbf{x}^{DD}_d$ and channel $(\hat{h}_r,\hat{\tau},\hat{\nu})$ using Eq.~(\ref{eq:DD_unfold});
    \State Cancel $\mathbf{y}_d^{sim}$ from $\mathbf{y}_{\text{uf}}^{DD}$ and update $\mathbf{y}_{\text{uf}}^{DD}$;
    \State Perform the MLE estimation to modified $\mathbf{y}_{\text{uf}}^{DD}$ to get a new estimation $(\hat{h}_r',\hat{\tau}',\hat{\nu}')$;
\If{$|\hat{\tau}'-\hat{\tau}| \le \epsilon$ \& $|\hat{\nu}'-\hat{\nu}| \le \epsilon$}
\State Update $(\hat{h}_r,\hat{\tau},\hat{\nu})$ and terminate;
\EndIf
\State Update $(\hat{h}_r,\hat{\tau},\hat{\nu})=(\hat{h}_r',\hat{\tau}',\hat{\nu}')$
\EndWhile 
\end{algorithmic}
\end{algorithm}

Although a careful design ensures that pilots are free from mutual interference, interference from data symbols is still unavoidable. Therefore, an iterative cancellation and detection algorithm is essential in this context. In this section, we describe a two-phase algorithm of iterative interference cancellation and detection. 

By properly selecting the parameters, we can generate an OTFS signal that the pilot symbols are mutually interference free from each other in the frequency domain even after the aliasing effect. This allows us to unfold the folded signal in frequency domain to recover the original pilot signal in TF domain before aliasing, and further convert to DD domain for the final range and velocity estimations. The relationship between the transmitted DD signal $\mathbf{x}^{DD}$ and captured unfolded DD domain signal $\mathbf{y}_{\text{uf}}^{DD}$ can be written as:
\begin{equation}\label{eq:DD_unfold}
\begin{array}{l}
\mathbf{y}_{\text{uf}}^{DD} = \sum\limits^{P}_{i=1} (\mathbf{F}_{N_{\nu}} \otimes \mathbf{G}_{rx})\mathbf{B}_{u}\mathbf{B}_{a} \Pi^{l_i}_{M_{\tau}N_{\nu}} \Delta_{M_{\tau}N_{\nu}}^{k_i} \\
\qquad \qquad \qquad \qquad \qquad \qquad \quad (\mathbf{F}^H_{N_{\nu}} \otimes \mathbf{G}_{tx}) \mathbf{x}^{DD},
\end{array}
\end{equation}
where $\mathbf{B}_{a} \in \mathbb{C}^{\frac{M_{\tau}}{\kappa}\times M_{\tau}}$ and $\mathbf{B}_{u} \in \mathbb{C}^{M_{\tau}\times \frac{M_{\tau}}{\kappa}}$ represent the aliasing matrix and unfolding matrix. The unfolding matrix \(\mathbf{B}_{u}\) is constructed by interpolating the aliased subcarriers with \(\mu - 1\) zero-inserted subcarriers between adjacent subcarriers, and by then sorting these aliased subcarriers according to their aliasing patterns.

Since the pilot is designed as an impulse sequence with a period of $\mu$ in the DD domain, the maximum unambiguous sensing range is reduced by a factor of $\mu$ when considering only the pilot signals. This phenomenon, known as the range ambiguity, arises from the radar's inability to uniquely determine the true distance to the target. This behavior is analogous to OFDM radar, where the maximum unambiguous range $r_{\text{max}}$ is inversely proportional to the subcarrier spacing~\cite{unambiguous1,SA} as $r_{\text{max}} = \frac{c}{2\Delta_f}$, where $c$ is the light speed and $\Delta_f$ is the subcarrier spacing. In our setup, pilots are periodically placed along the delay axis with fixed interval $\mu$ in the DD domain, resulting in the periodic activation of subcarriers in the TF domain, while other subcarriers remain inactive. This is equivalent to increasing the subcarrier spacing by a factor of $\mu$, which consequently degrades the maximum unambiguous sensing range to $\frac{c}{2\mu \Delta_f}$. However, in addition to the pilot symbols, leveraging the transmission data can provide supplementary information to eliminate the ambiguity. This approach is equivalent to utilizing the entire bandwidth without any idle subcarriers, ensuring that the system maintains both sensing resolution and maximum sensing range without degradation.

In the first phase of the algorithm, we aim to obtain a coarse estimation without range ambiguity. Initially, we perform a MLE on the unfolded signal $\mathbf{y}_{\text{uf}}^{DD}$ after Eq.~(\ref{eq:DD_unfold}) with the transmitted pilot signals $\mathbf{x}^{DD}_p$, while disregarding the transmission data. In this step, the aliased transmission data $\mathbf{x}^{DD}_d$ is treated entirely as an interference. By considering only the pilot signals in the estimation, the unambiguous range is reduced, which may lead to an inaccurate estimation. Subsequently, based on this initial estimation, we adjust the estimation by adding or subtracting the maximum unambiguous range to account for all potentially correct range values. We then iterate through these potential range values, applying virtual channels based on these values to the transmission data. For each potential channel, we calculate the Euclidean distance between the resulting signal and the received signal. This process helps eliminate the range ambiguity and yields an accurate estimation.

After obtaining a coarse estimation, the algorithm proceeds to the second phase, where we iteratively cancel the interference from the transmission data and refine the estimation of the target parameters. At each iteration, MLE is employed for channel coefficients, range and Doppler estimations $(\hat{h}_r,\hat{\tau},\hat{\nu})$. Then the estimations are iterated until convergence. The detailed algorithm is described in Algorithm~\ref{alg:cap}.

\section{Communication with Reduced Sampling Rate}
\label{sec:comms}

As discussed in the previous section, the reduced sampling rate can infer radar targets' ranges and velocities without compromising resolution and maximum sensing range. This is because, in a monostatic radar system, the transmitted signal is fully known to the transceiver, and the channel estimation requires much less information, which can be efficiently handled even with reduced bandwidth. However, for communication systems, the transmission capacity is directly constrained by the signal bandwidth, which is determined by the sampling rate. A reduced sampling rate leads to a proportional reduction in the data rate. To address this problem, we design a specific data embedding approach to embed the data.

\begin{figure}[!t]
 \centering
 \includegraphics[width=2.8in]{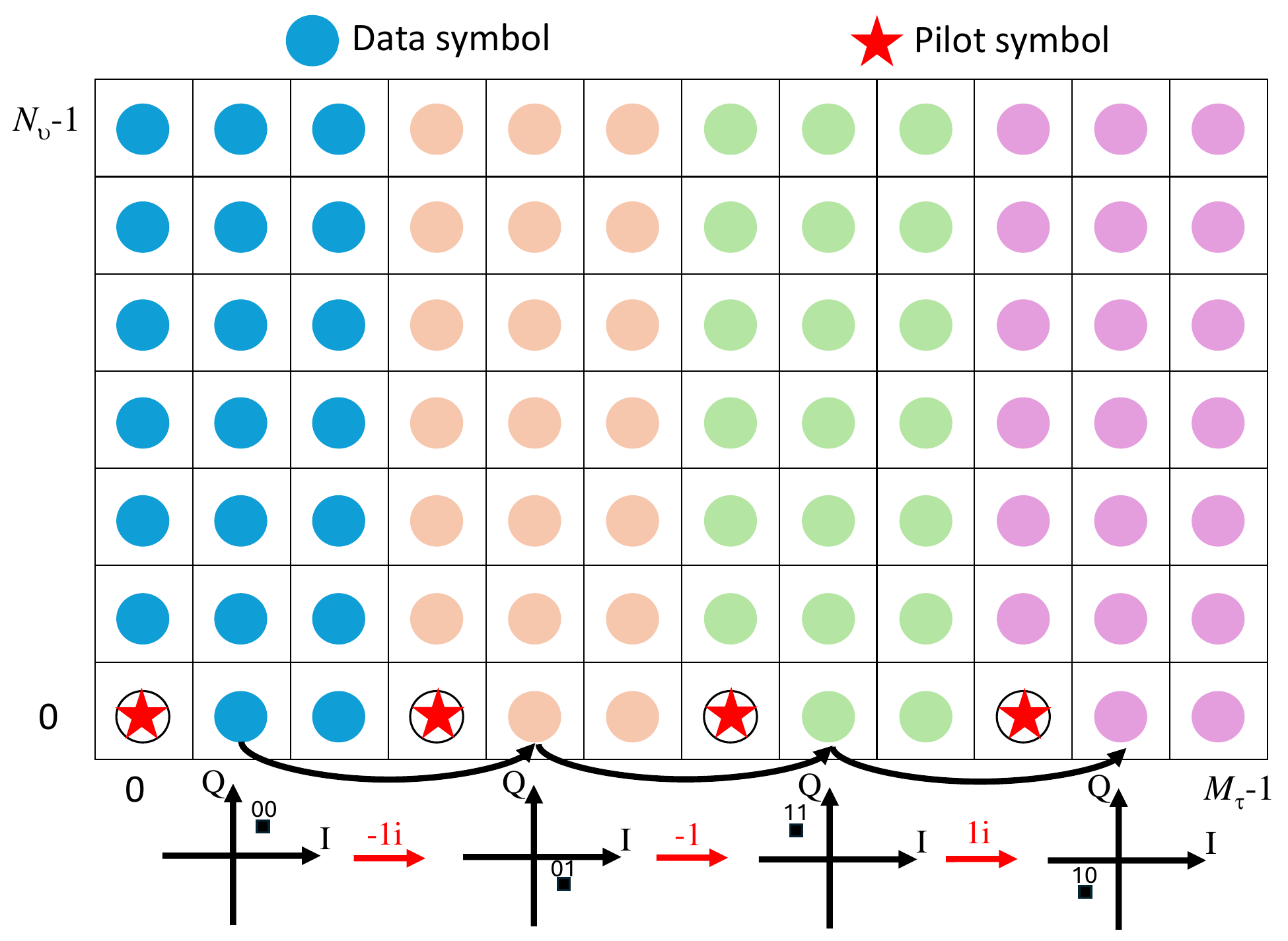}\\
 \caption{Code spreading in DD domain.}
 \label{fig:pilot_placement_comms}
\end{figure}

\subsection{Data Embedding}

To maintain compliance with the Shannon-Nyquist theorem, the transmitter must proportionally reduce its data transmission rate. Specifically, if the receiver down-samples the signal by a factor of $\frac{1}{\kappa}$ relative to the available bandwidth, then the transmitter can only allocate $\frac{1}{\kappa}$ of its resources to data embedding.

In our proposed scheme, we partition the DD domain resources along the delay axis into $\kappa$ slices of sub-bands, as illustrated in Figures~\ref{fig:pilot_placement} and \ref{fig:pilot_placement_comms}. Effective transmission data $\mathbf{X}_{e}^{DD} \in \mathbb{C}^{\frac{M_{\tau}}{\kappa}\times N_{\nu}}$ is embedded into the first slice, represented by the blue symbols in Fig.~\ref{fig:pilot_placement_comms}. This configuration ensures the maximum effective transmission data rate, complying with the Shannon-Nyquist theorem, when using a sampling rate reduced by a factor of $\kappa$ on the communication receiver side. To avoid degrading the maximum unambiguous sensing range, as discussed in the previous section, it is crucial not to leave the remaining DD resources empty. To address this, we need a spreading technique that distributes the effective transmission data across the entire domain. While in our scenario, traditional waveform-based spreading techniques such as direct sequence spread spectrum (DSSS), chirp spreading, or other windowing techniques are not feasible. This limitation arises because the receiver operates with downsampled sampling, making it challenging to despread the received signal directly. 

To enable effective spreading and despreading of signals at a reduced sampling rate on the receiver side without the need for additional analog hardware, we propose a spreading-code-based approach. A pair of transmitter and receiver maintain a sequence of length $\kappa$ for spreading purpose. Consider a multi-user scenario, where the sequence associated with the \( i \)-th transmitter and the \( j \)-th receiver is a Zadoff-Chu sequence denoted as \( S^{i,j} = \{S_1^{i,j},S_2^{i,j},\cdots,S_{\kappa}^{i,j}\} \). Zadoff-Chu sequences are complex-valued sequences with constant amplitude and ideal correlation properties, defined as
\begin{equation}\label{eq:zad_off}
   S^{i,j}_n = e^{-j\frac{\pi r^{i,j}n(n+1)}{\kappa}}, \quad n = 1, \dots, \kappa, 
\end{equation}
where \( r^{i,j} \) is the root index that determines the sequence and ensures its uniqueness for each transmitter-receiver pair. In the following, we use QPSK as an example, while this approach can be seamlessly extended to other modulation schemes with different modulation orders. After the effective transmission data $\mathbf{X}_{e}^{DD} $embedded onto the DD domain, it spreads out to the entire domain to generate the ultimate transmission data $\mathbf{X}^{DD} \in \mathbb{C}^{M_{\tau}\times N_{\nu}}$ using the corresponding code sequence:
\begin{equation}\label{eq:spreading_DD}
    \begin{aligned}
        \mathbf{X}^{DD} &= \mathbf{S}^{DD} \mathbf{X}_{e}^{DD} \\
        &= (S^{i,j}\otimes \mathbf{I}_{\frac{M_{\tau}}{\kappa}})\mathbf{X}_{e}^{DD},
    \end{aligned}
\end{equation}
where $\mathbf{I}_{\frac{M_{\tau}}{\kappa}}$ denotes the identity matrix of size $\frac{M_{\tau}}{\kappa}$. The vectorized form can be written as
\begin{equation}\label{eq:spreading_vec}
    \begin{aligned}
        \mathbf{x}^{DD} 
        &= (\mathbf{I}_{N_{\nu}} \otimes (S^{i,j}\otimes \mathbf{I}_{\frac{M_{\tau}
}{\kappa}})) \mathbf{x}_{e}^{DD},
    \end{aligned}
\end{equation}
where $\mathbf{x}^{DD}=vec(\mathbf{X}^{DD})$ and $\mathbf{x}_e^{DD}=vec(\mathbf{X}_e^{DD})$. Once the ultimate transmission data is ready, it is transmitted via the channel in Eq.~(\ref{eq:channelH}). Based on Equations (\ref{eq:IO}) and (\ref{eq:Gamma}), the input-output (IO) relationship between $\mathbf{x}_e^{DD}=vec(\mathbf{X}_e^{DD})$ and captured signal $\mathbf{y}^{DD}$ with full sampling rate system can be written as: \begin{equation}\label{eq:IO_full}
\begin{aligned}
\mathbf{y}^{DD} &= H\mathbf{x}^{DD} + \mathbf{w} \\
&= H (\mathbf{I}_N \otimes (S^{i,j}\otimes \mathbf{I}_{\frac{M}{\kappa}})) \mathbf{x}_{e}^{DD} + \mathbf{w}\\
&= G\mathbf{x}_{e}^{DD} + \mathbf{w}, 
\end{aligned}
\end{equation}
where $G$ denotes the DD channel matrix involving code spreading:
\begin{equation}\label{eq:Gamma_full}
\begin{array}{l}
G = \sum\limits^{P}\limits_{i=1} h_i (\mathbf{F}_{N_{\nu}} \otimes \mathbf{G}_{rx}) 
 \Pi^{l_i}_{M_{\tau}
N} \Delta_{M_{\tau}N_{\nu}}^{k_i} (\mathbf{F}^H_{N_{\nu}} \otimes \mathbf{G}_{tx}) \\ \qquad \qquad 
 \qquad \qquad \qquad \qquad (\mathbf{I}_{N_{\nu}} \otimes (S^{i,j}\otimes \mathbf{I}_{\frac{M_{\tau}
}{\kappa}})).
\end{array}
\end{equation}

The IO relationship with full sampling rate in Equations (\ref{eq:IO_full}) and (\ref{eq:Gamma_full}) can be further extended to the case when the receiver is downsampled by a factor of $\kappa$. The IO relationship between $\mathbf{x}_e^{DD}$ and the captured signal $\Tilde{\mathbf{y}}^{DD}$ with reduced sampling rate system can be written as
\begin{equation}\label{eq:IO_reduced}
\begin{aligned}
\Tilde{\mathbf{y}}^{DD} 
&= \Tilde{G}\mathbf{x}_{e}^{DD} + \mathbf{w}, 
\end{aligned}
\end{equation}
where $\Tilde{G}$ is the channel matrix of reduced sampling rate system\footnote{Notations with a tilde indicate counterparts in the reduced sampling rate system, relative to the full sampling rate system, unless stated otherwise.}:
\begin{equation}\label{eq:Gamma_reduced}
\begin{array}{l}
\Tilde{G} = \sum\limits^{P}\limits_{i=1} h_i (\mathbf{F}_{\frac{N_{\nu}}{\kappa}} \otimes \mathbf{G}_{rx}) \mathbf{D}_{\kappa}
 \Pi^{l_i}_{M_{\tau}
N} \Delta_{M_{\tau}N_{\nu}}^{k_i} (\mathbf{F}^H_{N_{\nu}} \otimes \mathbf{G}_{tx}) \\ \qquad \qquad 
 \qquad \qquad \qquad \qquad (\mathbf{I}_{N_{\nu}} \otimes (S^{i,j}\otimes \mathbf{I}_{\frac{M_{\tau}
}{\kappa}})),
\end{array}
\end{equation}
where $\mathbf{D}_{\kappa} \in \mathbb{C}^{\frac{M_{\tau}N_{\nu}}{\kappa}\times M_{\tau}N_{\nu}}$ denotes the downsampling matrix with rate $\kappa$:
\begin{equation}\label{eq:downsample_matrix}
\begin{array}{l}
\mathbf{D}_{\kappa} = \mathbf{I}_{\frac{M_{\tau}

}{\kappa}} \otimes \underbrace{[1, 0, \cdots, 0]}_{\text{$\kappa$ elements}}.
\end{array}
\end{equation}

Given the IO relationships established in Equations (\ref{eq:Gamma_reduced}) and (\ref{eq:downsample_matrix}), various equalization and decoding techniques, such as the Minimum Mean Square Error (MMSE) estimator, can be subsequently employed for effective data decoding.

\subsection{Synchronization with Reduced Sampling Rates}

In communication systems where the transmitter and receiver operate at mismatched sampling rates, synchronization poses significant challenges. This disparity complicates the accurate estimation of timing and frequency offsets, both of which are essential for maintaining communication integrity. Extensive research in signal processing has addressed synchronization challenges in systems with mismatched sampling rates~\cite{sync_mismatched1,sync_mismatched2,sync_mismatched3}, proposing various interpolation-based solutions to manage these issues. In our system, which operates at reduced sampling rates, we encounter synchronization difficulties due to decreased temporal and frequency resolution at the receiver. We have opted to implement a vanilla sinc interpolation-based approach. This choice is practical as systems with integer-reduced sampling rates are straightforward to analyze, and sinc interpolation effectively ensures sufficient synchronization accuracy for our communication needs, which are less stringent than those required in radar applications. Let the transmitted baseband signal be denoted by $x(t)$, at sampling rate $f_s$:

\begin{equation}
    x[n] = x(nT_s), \quad n \in \mathbb{Z}, \quad T_s = \frac{1}{f_s}.
\end{equation}

At the receiver, the signal is sampled at $f'_s = \frac{f_s}{\kappa}$:

\begin{equation}
    \tilde{y}[m] = x(mT'_s) + w[m], \quad m \in \mathbb{Z}, \quad T'_s = \frac{1}{f'_s},
\end{equation}
where $w[m]$ represents additive white Gaussian noise (AWGN). Considering both timing and frequency offsets, the received signal can be expressed as

\begin{equation}
    \tilde{y}[m] = x(mT'_s - \tau) e^{j2\pi \Delta f mT'_s} + w[m],
\end{equation}
where $\tau$ and $\Delta f$ denote the timing and frequency offset separately. Our objective is to accurately estimate and compensate for $\tau$ and $\Delta f$ despite the reduced sampling rate. 

\subsubsection{Time Offset Estimation}
We utilize cross correlation with interpolation to achieve fractional sampling synchronization. The interpolation process aims to reconstruct the original full sampling rate signal $y[k]$ from the low-rate samples $\tilde{y}[m]$. For practical implementation, a sinc interpolation filter can be used to minimize the aliasing~\cite{sinc_interpolation}:
\begin{equation}
    y[k] = \sum_{m=-\infty}^{\infty} \tilde{y}[m] \cdot \text{sinc}\left( \frac{k - m\kappa}{\kappa} \right),
\end{equation}
where \(\text{sinc}(x) = \frac{\sin(\pi x)}{\pi x}\). Here, the interpolation kernel rescales the downsampled grid by a factor of \(\kappa\), effectively reconstructing the samples at the original rate. Thus, the interpolated signal \(y[k]\) aligns closely with the transmitter’s sampling instances. In practice, this sum is truncated and a finite-length filter approximates the ideal sinc function. Despite the truncation, the dominant contribution typically comes from terms near \( m \approx \frac{k}{\kappa} \), making the truncated signal a good approximation of \( y[k] \).

With \(y[k]\) approximating the full-rate received sequence, we can estimate the timing offset \(\tau\) by correlating \(y[k]\) with the known transmitted reference sequence \( x[n] \). Define the cross-correlation as
\begin{equation}
  R_{yx}[\ell] = \sum_{l} y[l]\, x^{*}[l - \ell],  
\end{equation}
where \((\cdot)^{*}\) denotes complex conjugation. The value of \(\hat{\ell}\) that maximizes \(|R_{yx}[\ell]|\) provides an estimate of the timing offset. More explicitly, we have $\hat{\tau} = \hat{\ell}T_s$. 

\subsubsection{Frequency Offset Estimation}

After we have estimated and compensated the timing offset \(\hat{\tau}\), the next step is to estimate the frequency offset \(\Delta f\). One common approach is to use an MLE. Assume that the timing offset has been corrected; therefore the received samples can be modeled as
\begin{equation}
 y[m] \approx x(mT_s' - \hat{\tau}) e^{j2\pi \Delta f m T_s'} + w[m].   
\end{equation}
The MLE for \(\Delta f\) is obtained by maximizing the following likelihood function:
\begin{equation}
   \hat{\Delta}f = \arg \max_{\Delta f} \sum_{m} y[m]x^{*}(mT_s' - \hat{\tau}) e^{-j2\pi \Delta f m T_s'}. 
\end{equation}

To find the maximum, we take the derivative with respect to \(\Delta f\), set it to zero, and solve for \(\Delta f\). Differentiating and rearranging terms, we have
\begin{equation}
 \hat{\Delta}f = \frac{1}{2\pi}\frac{\sum_{m} m T_s' \,\text{Im}\{ y[m] x^{*}(mT_s' - \hat{\tau}) \}}{\sum_{m} |y[m]x^{*}(mT_s' - \hat{\tau})|^2},   
\end{equation}
where \(\text{Im}\{\cdot\}\) denotes the imaginary parts.

\subsection{Frame Structure}

Since OTFS is specifically designed for high-mobility applications, where channel aging is inherently severe due to rapid temporal variations in the channel state, this necessitates frequent updates to channel estimation to maintain optimal channel estimation. We assume a communication channel is represented by the complex vector \( \bm{h}_{n}\), where \( n \) denotes the block index. For simplicity, the channel aging is modeled as a frequency-flat fading, whose dynamics are expressed as~\cite{channel_aging_model}
\begin{equation}
    \bm{h}_{n} = \rho_q \bm{h}_{n-1} + \sqrt{1 - \rho_q^2} \bm{e}_{q,n},
\end{equation}
where \( \rho_q \) denotes temporal correlation coefficient dependent on the Doppler shift \( f_q \), modeled as \( \rho_q = J_0(2 \pi f_q T) \) using the Jakes' model~\cite{Jakes_model}. Here, \( J_0(\cdot) \) is the zeroth-order Bessel function of the first kind. \( \bm{e}_{q,n} \sim \mathcal{CN}(0, \beta_q \bm{I}_{L_t}) \) denotes the complex Gaussian noise independent of \( \bm{h}_{n-1} \), representing new randomness in the channel. \( \beta_q \) is the large-scale fading coefficient. It is demonstrated in~\cite{channel_aging_impact} that channel aging can have a severe impact to communication performance under high mobility scenario.

Additionally, due to the periodic nature of pilot placement of the ISAC signal along delay axis in the DD domain, as illustrated in Fig.~\ref{fig:pilot_placement} and Fig.~\ref{fig:overall performance of waveform recovery}(a), channel estimation is highly susceptible to delay ambiguity. To address this issue and ensure accurate channel delay estimation, it is essential to include a dedicated channel estimation block capable of resolving absolute channel parameters without delay ambiguity.

\begin{figure}[!t]
 \centering
 \includegraphics[width=2.5in]{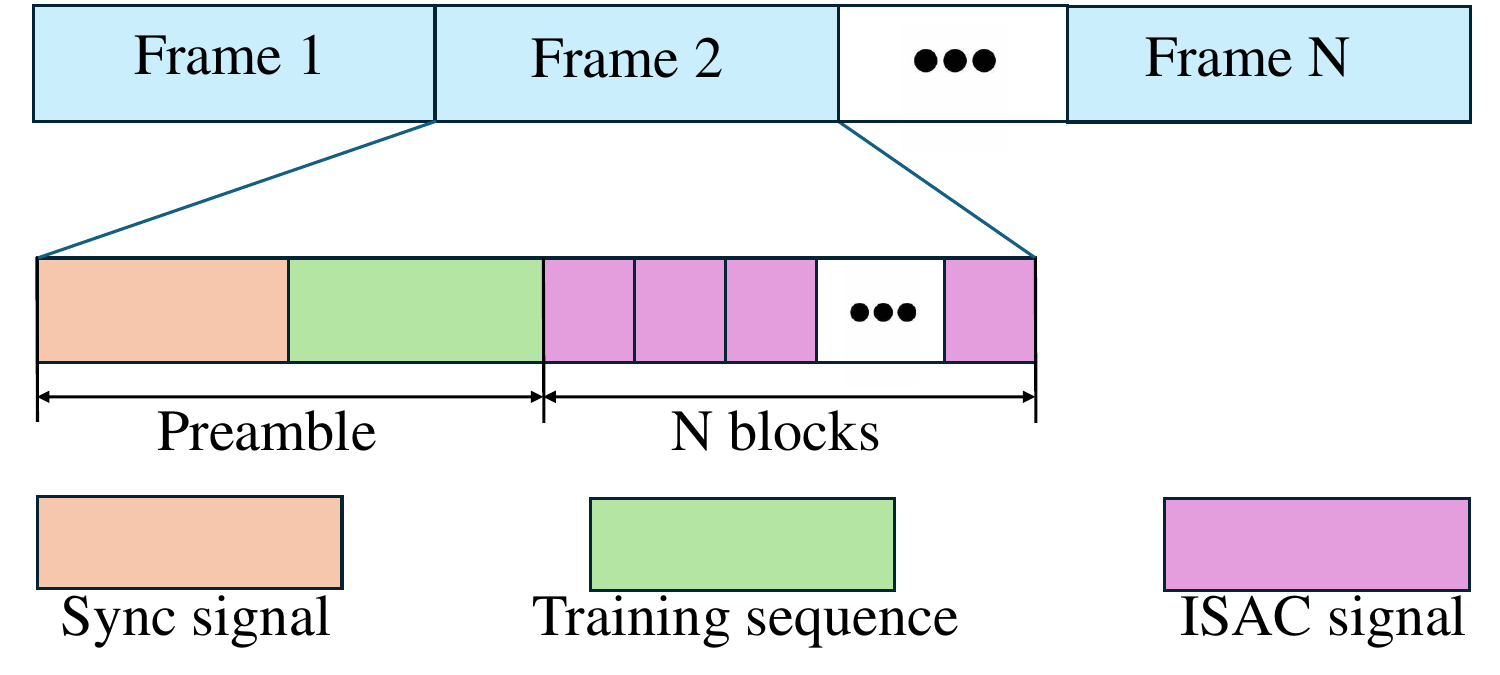}\\
 \caption{Illustration of the frame structure.}
 \label{fig:frame_structure}
\end{figure}

Based on the aforementioned requirements, we design the frame structure shown in Fig.~\ref{fig:frame_structure}. Each frame includes a preamble and multiple ISAC signal blocks. The preamble, containing a chirp-based synchronization signal and training sequence, enables frame-based channel estimation. By keeping the frame length within the coherence time, channel information (e.g., the number of paths, delays, Doppler shifts) remains constant throughout the frame. Each ISAC block (Fig.~\ref{fig:pilot_placement}) embeds pilot and data symbols. The training sequence provides slow-varying channel information, while block-by-block updates refine channel coefficients, minimizing performance degradation in high-mobility scenarios. Absolute delay estimates in the training sequence also resolve delay ambiguities.

\subsection{Iterative Channel Estimation and Data Decoding}

Since we utilize the superimposed pilot scheme to improve the communication efficiancy, an iterative channel estimation and data decoding algorithm~\cite{superimposed_OTFS} is necessary. Contrary to systems that operate at full sampling rates, our reduced sampling rate system necessitates specific adaptations of the iterative algorithm. Specifically, these modifications involve tailored channel coefficient estimation to accommodate our novel pilot placement, and a data detector based on the IO relationship specific to the reduced sampling rate system, as derived in Eq.~(\ref{eq:Gamma_reduced}). The detailed algorithm is as follows, First, an initial channel estimation is achieved based on the training sequence, delivering the number of channel paths $P$, channel delay taps $l_{1,2,\cdots,P}$ and channel Doppler taps $k_{1,2,\cdots,P}$ using Algorithm~\ref{alg:cap}. Then the reduced sampling rate ISAC signal is unfolded in the DD domain using Eq.~(\ref{eq:DD_unfold}) to obtain $\mathbf{y}_{\text{uf}}^{DD}$. Then a coarse channel coefficient estimation is implemented for each channel path based on $\mathbf{y}_{\text{uf}}^{DD}$. In our system, since we have multiple pilots instead of a tradition case with a single peak pilot, an averaging operation can be applied to improve the channel coefficient estimations. The channel coefficient estimation of the $p$-th path can be written as
\begin{equation}\label{eq:channel_coef_estimation}
  \hat{h}_p\left[ (l - l_p)_{M_{\tau}
},(k - k_p)_{N_{\nu}} \right] = \frac{1}{\mu} \sum_{j=0}^{\mu-1} \frac{\mathbf{y}_{\text{uf}}^{DD}[l+\frac{jM_{\tau}}{\mu},k]}{x_p},  
\end{equation}
where $x_p$ denotes the pilot value. After obtaining the channel estimation, we now use the MMSE to detect data by
using data-aided channel estimate. Given this vectorized input-output relation in OTFS in Eq.~(\ref{eq:Gamma_reduced}), MMSE can be utilized to estimate the transmitted effective data $\mathbf{X}_e^{DD}$: 

\begin{equation}\label{eq:MMSE}
\begin{array}{l}
\hat{\mathbf{x}}_e^{DD} = (\Tilde{G}^H \Tilde{G} +\sigma^2\mathbf{I}_{\frac{M_{\tau}N_{\nu}}{\kappa}})^{-1}\Tilde{G}^H\Tilde{\mathbf{y}}^{DD}
\end{array}
\end{equation}

After OTFS symbol detection, we have the symbols given by $\mathbf{X}_e^{DD}$. Then we are able to cancel out the interference induced by the data symbols based on the estimated channel and the obtained symbols, given by
\begin{equation}\label{eq:cancellation}
    \Tilde{\mathbf{y}}'^{DD} = \Tilde{\mathbf{y}}^{DD} - \Tilde{G}\hat{\mathbf{x}}_e^{DD}.
\end{equation}

If the interference is perfectly cancelled, then the residual term $\Tilde{\mathbf{y}}'^{DD}$ contains only the pilot information and the noise. Nevertheless, due to the impact of noise, cancellation is generally imperfect. Therefore, we employ a scheme that iteratively performs data detection and channel estimation. In the \(r\)-th iteration, the channel estimate \(\hat{h}^r\) is obtained using the signal after interference cancellation from the previous iteration. Subsequently, the data symbols \(\hat{\mathbf{x}}_e^{r}\) are detected based on the estimated channel, \(\hat{h}^r\). Both \(\hat{\mathbf{x}}_e^{r}\) and \(\hat{h}^r\) are then utilized for the interference cancellation, generating the interference-canceled signal for use in the \((r+1)\)-th iteration. After multiple iterations, the channel estimates and the detected data symbols are refined and finalized. The iterative process is terminated when the performance improvement from additional iterations becomes negligible.

\begin{algorithm}[!t]
\caption{The iterative channel estimation and data detection algorithm}\label{alg:data decoding}
\begin{algorithmic}[1]
\State \textbf{Input:} Convergence error $\epsilon$, received signal $\Tilde{\mathbf{y}}^{DD}$, channel delays taps $l_{1,2,\cdots,P}$, channel Doppler taps $k_{1,2,\cdots,P}$, number of multipaths $P$.

\State \textbf{Output:} Decoded data $\hat{\mathbf{x}}_e^{DD}$;
\State \textbf{Initialize:} Counter $r = 1$;
\State Unfold the received signal $\Tilde{\mathbf{y}}^{DD}$ using Eq.~(\ref{eq:DD_unfold}) to get $\mathbf{y}_{\text{uf}}^{DD}$;
\While{not terminate}
\For{$i = 1\cdots P$}
\State Get the channel coefficient $\hat{h}^r_i$ using Eq.~(\ref{eq:channel_coef_estimation}) based on $\mathbf{y}_{\text{uf}}^{DD}$;
\EndFor
\State Construct the channel matrix $\Tilde{G}$ using Eq.~(\ref{eq:Gamma_reduced}) based on $\hat{h}^r_i$,$\tau$,$\mu$;
\State Use MMSE to decode data $\hat{\mathbf{x}}_e^{r}$ using Eq.~(\ref{eq:MMSE});
\State Cancel out the interference based on $\Tilde{G}$ and $\hat{\mathbf{x}}_e^{r}$ using Eq.~(\ref{eq:cancellation}) and get the canceled signal $\Tilde{\mathbf{y}}'^{DD}$;
\State Update the unfolded signal $\mathbf{y}_{\text{uf}}^{DD}$ based on $\Tilde{\mathbf{y}}'^{DD}$;
\If{$|\hat{h}^{r-1}-\hat{h}^{r}| \le \epsilon$ \& $\hat{\mathbf{x}}_e^{r-1}==\hat{\mathbf{x}}_e^{r}$}
\State $\hat{\mathbf{x}}_e^{DD} = \hat{\mathbf{x}}_e^{r}$ and terminate;
\EndIf
\State $r=r+1$
\EndWhile 
\end{algorithmic}
\end{algorithm}

\begin{figure*}[!t]
    \centering
    \subfigure[Ground truth of the targets' ranges and velocities.]
    {
        \label{fig:radar_sim1}
        \includegraphics[width=2.0in]{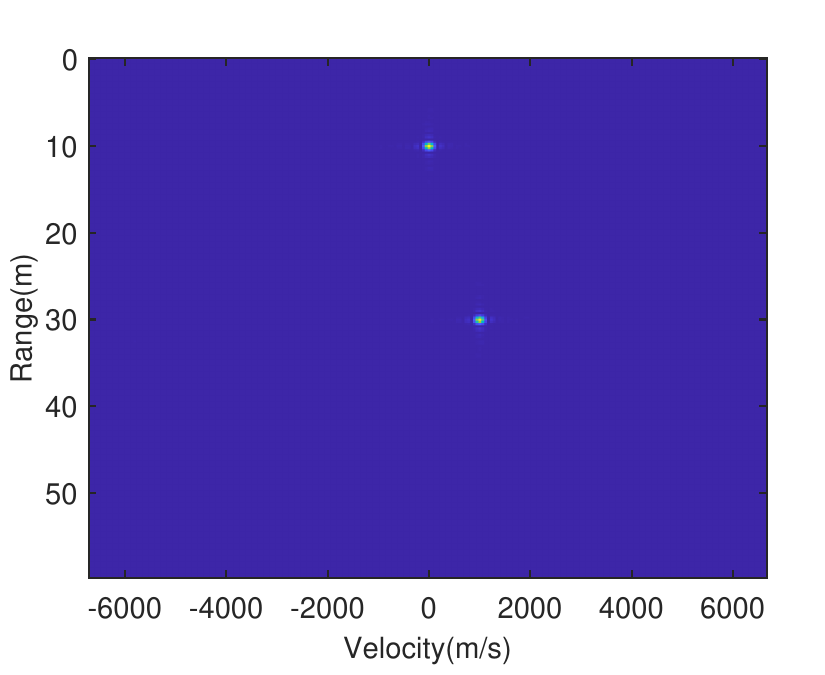}
    }
    \hspace{0.01\linewidth}
    \subfigure[Estimation result with only pilot.]
    {
        \label{fig:radar_sim2}
        \includegraphics[width=2.0in]{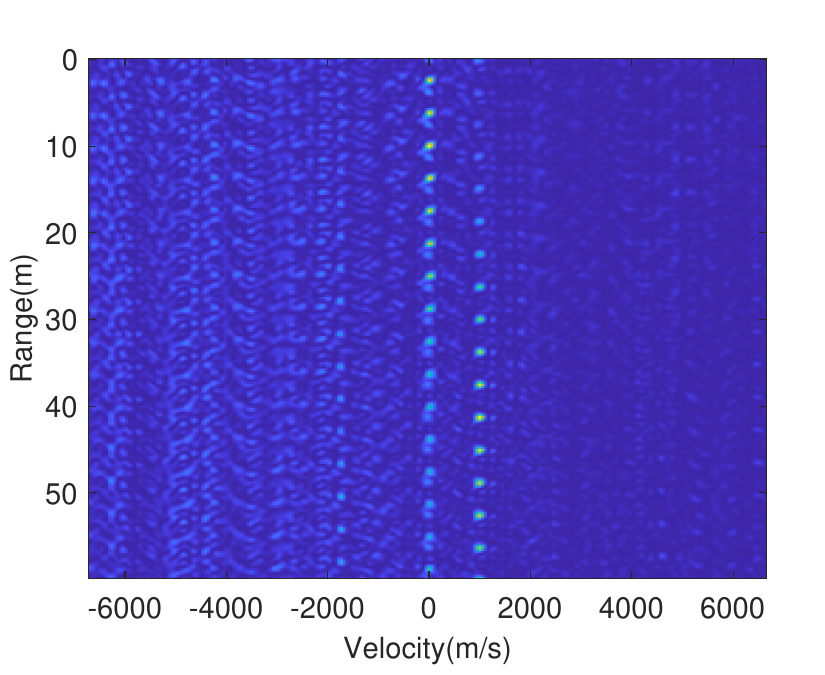}
    }
    \hspace{0.01\linewidth}
    \subfigure[Estimation result with iterative detection.]
    {
        \label{fig:radar_sim3}
        \includegraphics[width=2.0in]{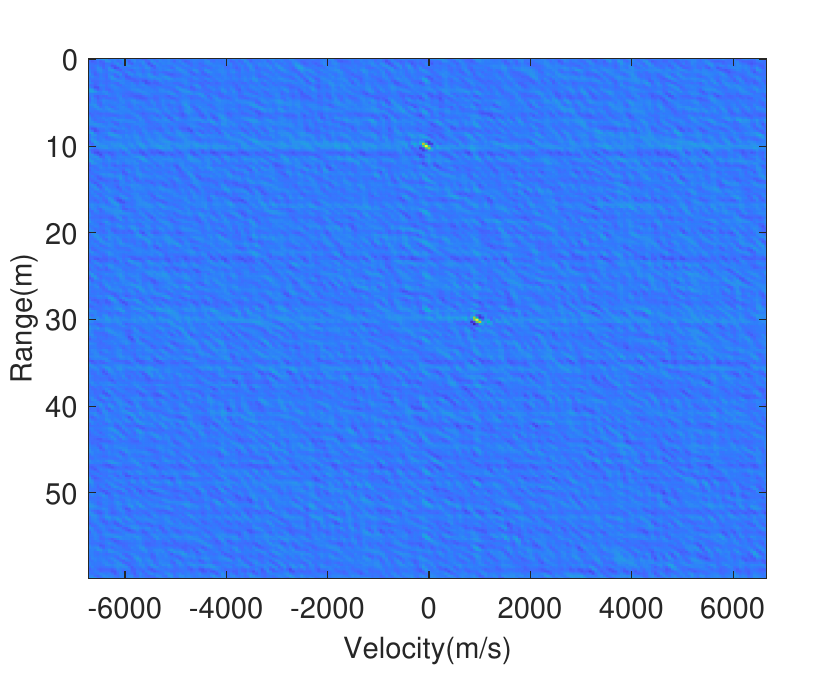}
    }
    \hspace{0.01\linewidth}
\caption{Simulation results of the range-velocity profile for two targets with distinct ranges and velocities.}
\label{fig:radar_sim}
\end{figure*}

\begin{figure}[!t]
   \centering
   \subfigure[Estimation result with two targets located within range resolution.]
   {
       \label{fig:radar_sim_resolution1}
        \includegraphics[width=0.44\columnwidth]{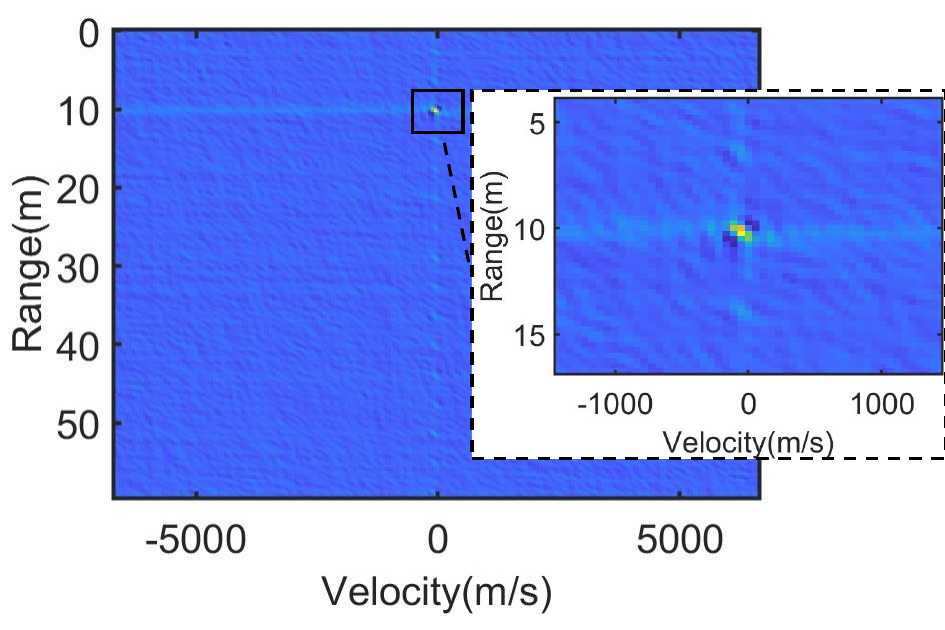}
   }
   \hspace{0.01\linewidth}
   \subfigure[Estimation result with two targets located beyond range resolution.]
   {
      \label{fig:radar_sim_resolution2}
        \includegraphics[width=0.44\columnwidth]{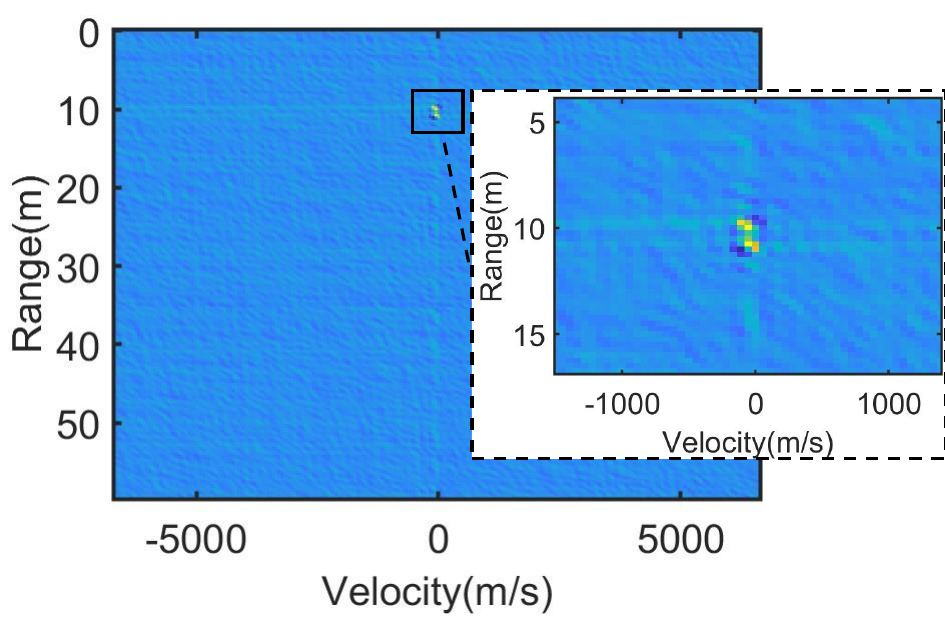}
   }
   \hspace{0.01\linewidth}
\caption{Simulation results of the range-velocity profile for two closely located targets.}
\label{fig:radar_sim_resolution}
\end{figure}

\section{Performance Analysis}
\label{sec:properties}

In this section, we compare of the performance of our reduced sampling rate system with the full sampling rate system, in the aspects of range resolution, maximum sensing range, velocity resolution, channel estimation performance, and communication performance.

\subsection{Effects on Radar Parameter Estimation}
The range resolution of a radar system is determined by its bandwidth. In our design, the pilot scheme is configured to occupy the same virtual bandwidth as that of the full-sampling-rate system, ensuring that the range resolution remains unaffected. Additionally, the maximum unambiguous sensing range increases with the number of subcarriers. By leveraging the data-aided estimation and distributing communication across the entire bandwidth, our system achieves the maximum sensing range without performance degradation or ghost target estimation. Furthermore, the reduced sampling rate system retains the same maximum velocity and velocity resolution capabilities as the full-sampling-rate system.

\subsection{Effects on Channel Estimation}\label{sec:effects_channel}
In the superimposed scheme, the interference from the data to the pilot is unavoidable. In the full sampling rate system, the interference term from data can be expressed as
\begin{equation}\label{eq:interference}
I_{k,l} = \sum_{i \in \mathcal{Q}} h_i \mathbf{X}_d^{DD}\left[ (l - l_{i})_{M_{\tau}
},(k - k_{i})_{N_{\nu}} \right] e^{-j 2 \pi \frac{k_{i} l_{i}}{N_{\nu}M_{\tau}
}}, 
\end{equation}
where $\mathcal{Q}$ denotes the set containing indices of all data symbols contributed to sample $\mathbf{Y}^{DD}[k, l]$. In particular, the pilot is interfered by $P$ data symbols, therefore $|\mathcal{Q}_{k,l}| = P$. Based on Eq.~(\ref{eq:interference}) and with the assumption that the channel coefficients are independent for different paths $\mathbb{E}\{h_i h_{i'}^*\} = 0$, the interference energy $\mathbb{E}_{\text{I}}=\mathbb{E}\left\{|I_{k,l}|^2\right\}$ can be expressed as~\cite{superimposed_OTFS}
\begin{equation}\label{eq:interference energy}
\begin{aligned}
\mathbb{E}_{\text{I}} &  = \sum_{i \in \mathcal{Q}_{k,l}} \mathbb{E}\left\{|h_i|^2\right\} E_s,
\end{aligned}
\end{equation}
where $E_s=\mathbb{E}\{|\mathbf{X}_d^{DD}[k,l]|^2\}$ denotes the average data symbol energy. Considering the normalized channel power gains, \(\sum_{i=1}^P \mathbb{E}\left\{|h_i|^2\right\} = 1\), the energy of the interference term is established as \(\mathbb{E}_{\text{I}} = E_s\). Consequently, the SINR for the pilot signal can be expressed as
\begin{equation}
    \text{SINR}_p = \frac{E_p}{N_0+E_s}.
\end{equation}
In the reduced sampling system, where the alisaing effect caused by downsampling will introduce further interference. As illustrated in Fig.~\ref{fig:aliasing}, the signal is folded into a signal band after aliasing, therefore the data and noises from different sub-band are summed together. Therefore the powers of interference and noises are $\mu$ times the full sampling system. Consequently, we have the SINR for reduced sampling rate system
\begin{equation}
    \widetilde{\text{SINR}}_p = \frac{E_p}{\mu(N_0+E_s)}.
\end{equation}

\subsection{Effects on Communication Performance}\label{sec:effects_comms}
When considering the SINR of the communication data, pilot signals are treated as interference. Following a similar analysis as in the Section~\ref{sec:effects_channel}, the SINR of the full sampling rate system is expressed as
\begin{equation}
    \text{SINR}_d = \frac{E_s}{N_0+E_p}.
\end{equation}
For the reduced sampling rate system, a known code sequence is employed to spread the data across the entire domain. Consequently, the data symbols folded from other sub-bands can be effectively utilized as communication data rather than being treated as interference. In this case, the SINR of the reduced sampling rate system remains unaffected.

\subsection{Effects on PAPR}
Although downsampling at the receiver does not directly affect the peak-to-average power ratio (PAPR), which depends solely on the transmitted signal, it introduces additional noise and interference caused by the aliasing effect, as discussed in Section~\ref{sec:effects_channel}. This degradation in signal quality can negatively impact the performance of channel estimation in a reduced sampling rate system. To ensure accurate channel estimation, increasing the pilot power is recommended. While According to~\cite{PAPR_analysis}, there is an upper bound of PAPR for transmitted signal with rectangle pulse shape:
\begin{equation}
  \begin{aligned} 
\text{PAPR}_{\text{max}} = \frac{N_{\nu} \max_{k,l} |\mathbf{X}_d^{DD}[l,k]|^2}{\mathbb{E}(|\mathbf{X}_d^{DD}[l,k]|^2)} \cdot
\end{aligned}  
\end{equation}

Increasing the pilot power inevitably raises the PAPR, which can lead to signal distortion if the signal peaks exceed the amplifier's limit. However, thanks to the design of our frame structure, an initial large-scale channel estimation is performed using a dedicated training sequence, thus providing a reliable estimation of channel delays and Doppler taps. Subsequently, by leveraging the ISAC signal, frequent updates are achieved to ensure that channel variations remain minimal across consecutive blocks. This approach relaxes the strict requirement for high pilot power.

\begin{table}[!b]
\centering
\caption{Detailed parameter settings}
\label{my-label}
\begin{tabular}{@{}l@{\hspace{10em}}c@{}}
\toprule
\textbf{Parameters}      & \textbf{Values}   \\ \midrule
Sampling rate at the transmitter ($f_s$)       & 200~\!MHz            \\
Downsample rate ($\kappa$)    & 16\\
Sampling rate at the receivers ($f_s'$)       & 12.5~\!MHz            \\
Central frequency ($f_c$) & 28~\!GHz\\
$\#$  of delay bins ($M_{\tau}$)          & 80           \\
$\#$  of Doppler bins ($N_{\nu}$)          & 80     
      \\ 
$\#$  of pilots (Np)       & 16      \\ 
Subcarrier spacing($\Delta_f$)      & 2.5~\!MHz             \\
\bottomrule
\end{tabular}
\label{tab:signal_settings}
\end{table}

\section{Simulation and Experimental Results}\label{sec:results}

In this section, we first numerically validate the algorithm and the performance of the proposed designs using simulation results, then demonstrate these findings using our hardware ISAC testbed.

\subsection{Simulation Results}

In this study, we explore an OTFS system characterized by
specific parameters: the number of delay bins ($M_{\tau} = 80$), Doppler
bins ($N_{\nu} = 80$), carrier frequency ($28~$GHz), and subcarrier
spacing ($2.5~$MHz). The system employs a rectangular pulse
shaping method. The downsampling rates at both the radar and communication receivers are 16. Both the simulation and practical experiments were conducted using the aforementioned
settings unless otherwise noted. The parameters of OTFS system are outlined in Table \ref{tab:signal_settings}. Under this setting, the range resolution ($\Delta_{r}$) and velocity resoultion ($\Delta_{\nu}$) are calculated as:
\begin{align}
\label{eq:resolution}
       \Delta_{r} &= \frac{c}{2\Delta_f M_{\tau}} \approx 0.75m\\
   \Delta_{\nu} &= \frac{c\Delta_f}{2f_c N_{\nu}} \approx 523m/s
\end{align}

\subsubsection{Radar Estimation Performance}
In the simulation, we artificially control the locations and velocities of the two targets. Given the large Doppler resolution in OTFS settings, distinguishing between the two targets along the Doppler axis is challenging. We assign one target an artificially high velocity, which, although unrealistic in practical scenarios, enhances the visualization of the effects of the proposed algorithms.

In the first experiment, we simulate two targets located at 10 meters and 30 meters with velocities of 0 m/s and 1000 m/s, respectively. The range-velocity profile is depicted in Fig.~\ref{fig:radar_sim1}. After performing a coarse estimation with only the unfolded pilot symbol, the result in Fig.~\ref{fig:radar_sim2} shows ambiguity in the range estimations. This range ambiguity arises from the reduced maximum sensing range, leading to a wrap-around effect that prevents true range estimation. However, the results obtained using our proposed algorithm, illustrated in Fig. \ref{fig:radar_sim3}, show that the ambiguity is eliminated, and the estimation closely aligns with the true locations and velocities of the targets. This demonstrates the effectiveness of our algorithm in achieving accurate estimations with a reduced sampling rate, without compromising the maximum sensing range.

In the second experiment, we set the targets close to each other, both being stationary. In Fig.~\ref{fig:radar_sim_resolution1}, the distance between the two targets is slightly below the range resolution given by Eq.~(\ref{eq:resolution}), specifically 0.7 meters. The figure shows a single peak, indicating that the two targets are indistinguishable. In Fig.~\ref{fig:radar_sim_resolution2}, we adjust the distance between the targets to slightly above the range resolution, at 0.8 meters. This adjustment results in two distinct peaks in the range-velocity profile, clearly distinguishing the two targets. These results validate the effectiveness of our algorithm in achieving accurate estimation using a reduced sampling rate without sacrificing the range resolution.

\begin{figure}[!t]
 \centering
 \includegraphics[width=2.6in]{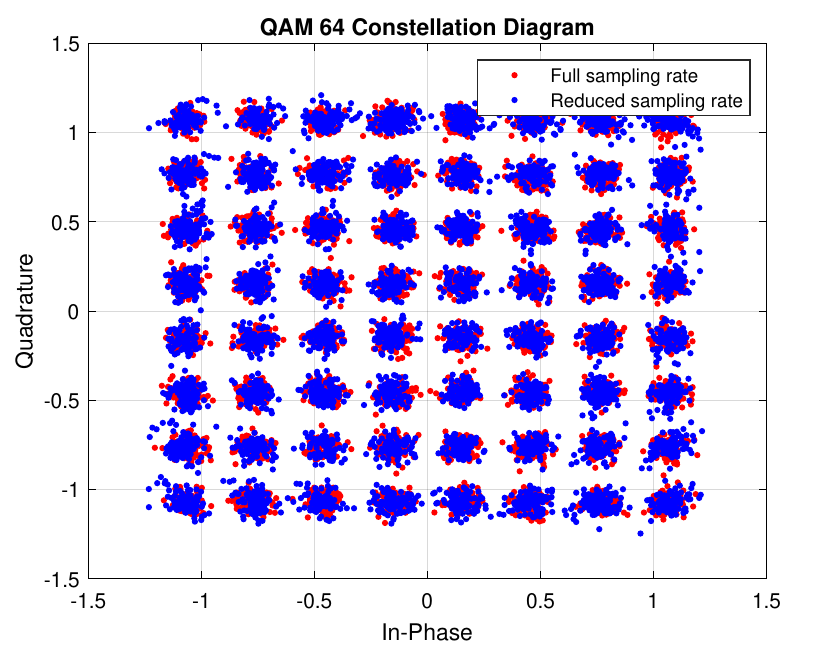}\\
 \caption{QAM 64 constellation diagram.}
 \label{fig:constellation}
\end{figure}

\begin{figure*}[!h]
    \centering
    \subfigure[BER of different modulation order with perfect synchronization and channel knowledge.]
    {
        \label{fig:BER_SNR_sim1}
        \includegraphics[width=2.1in]{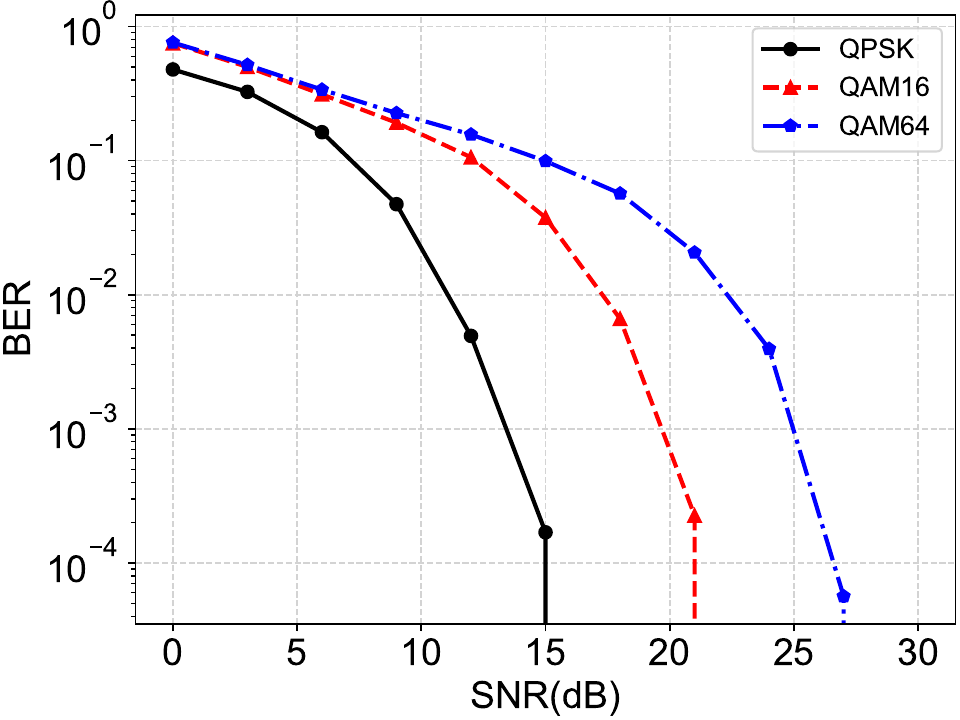}
    }
    \hspace{0.01\linewidth}
    \subfigure[BER of different modulation order with perfect synchronization and imperfect channel knowledge.]
    {
        \label{fig:BER_SNR_sim2}
        \includegraphics[width=2.1in]{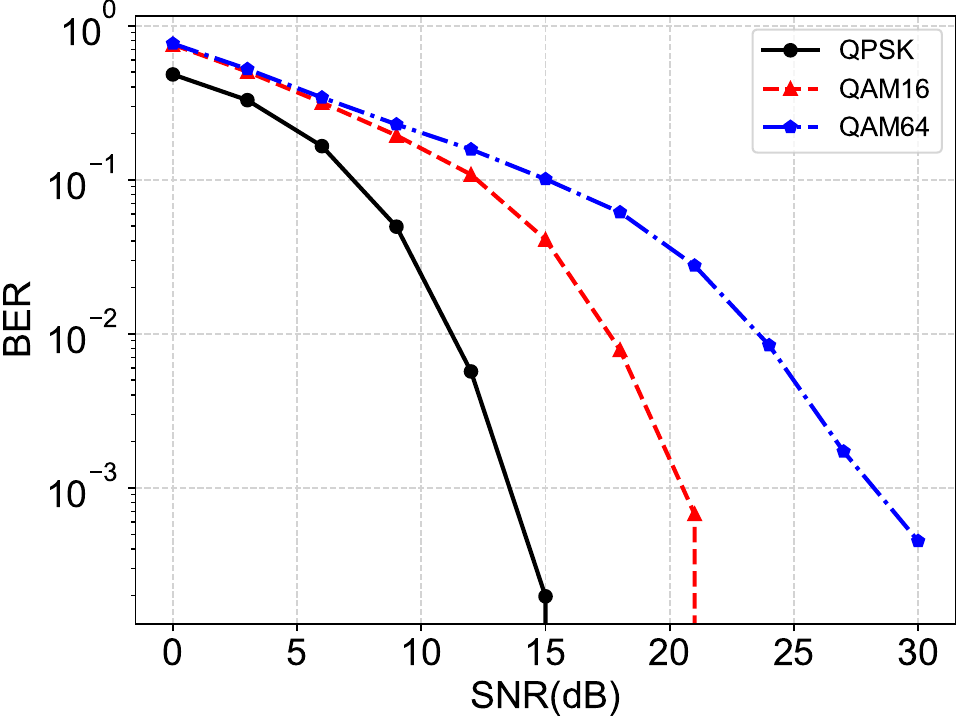}
    }
    \hspace{0.01\linewidth}
    \subfigure[BER of different modulation order with imperfect synchronization and imperfect channel knowledge.]
    {
        \label{fig:BER_SNR_sim3}
        \includegraphics[width=2.1in]{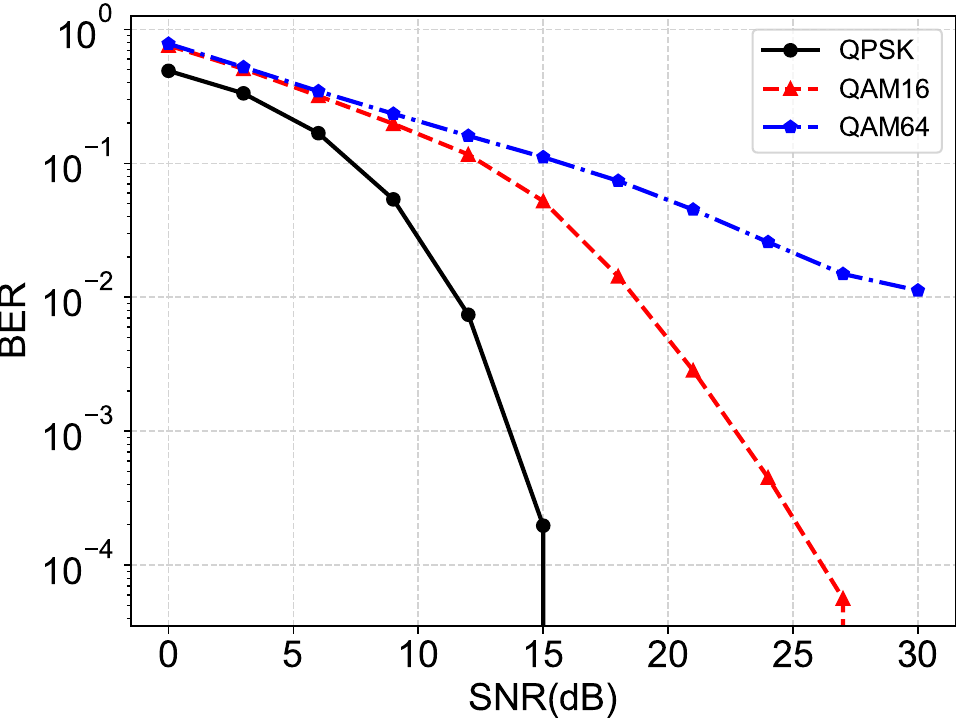}
    }
    \hspace{0.01\linewidth}
\caption{Simulation results of communication perforance with different modulation orders.}
\label{fig:BER_SNR_sim}
\end{figure*}

\subsubsection{Communication Performance}

We further simulate and analyze the communication performance. The simulation employed a Rician channel model, accompanied by a randomized channel impulse response (CIR) generation. Specifically, we simulated $1 \times 10^4$ distinct channel realizations for each signal-to-noise ratio (SNR) level within the Monte Carlo analysis framework. To evaluate the system performance, we considered three modulation schemes, including QPSK, 16-QAM, and 64-QAM.

In the first simulation, we compared the performance of the original full-bandwidth OTFS communication system with that of a reduced-sampling-rate system. The evaluation was conducted using 64-QAM modulation under a fixed SNR of 35 dB, assuming perfect channel knowledge. The resulting constellation diagrams, depicted in Fig.~\ref{fig:constellation}, demonstrate that the communication performance of the reduced-sampling system closely matches that of the full-sampling system. This result validates the theoretical analysis presented in Section ~\ref{sec:effects_comms}.

In the second simulation, we evaluated the performance of the reduced sampling rate system under three different channel assumptions: 1) with integer delay and Doppler shifts, and perfect channel knowledge; 2) with integer delay and Doppler shifts, and imperfect channel knowledge; 3) with fractional delay and Doppler shifts, and imperfect channel knowledge. As shown in Fig.~\ref{fig:BER_SNR_sim}, the BER performance for different modulation orders as the SNR increases from 0 to 30 dB under these three channel conditions is presented. In scenario (a), QPSK achieves the lowest BER, reaching \(10^{-4}\) at approximately 15 dB, while 16-QAM and 64-QAM achieve similar BER values at higher SNRs of 20 dB and 25 dB, respectively. In scenario (b), the BER degrades due to imperfect channel knowledge, with QPSK maintaining better noise tolerance and requiring around 17 dB to achieve \(10^{-4}\) BER, compared to 22 dB for 16-QAM and 28 dB for 64-QAM. Scenario (c) introduces the most severe degradation due to both imperfect synchronization and channel knowledge. Under these conditions, QPSK still performs relatively well, requiring 20 dB to reach a \(10^{-4}\) BER, whereas 16-QAM and 64-QAM exhibit significant a BER degradation, requiring SNR values exceeding 25 dB and 30 dB, respectively.

\subsection{Experimental Results}

\begin{figure}[!htb]
   \centering
   \subfigure[Communication TX and monostatic radar.]
   {
       \includegraphics[width=0.45\columnwidth]{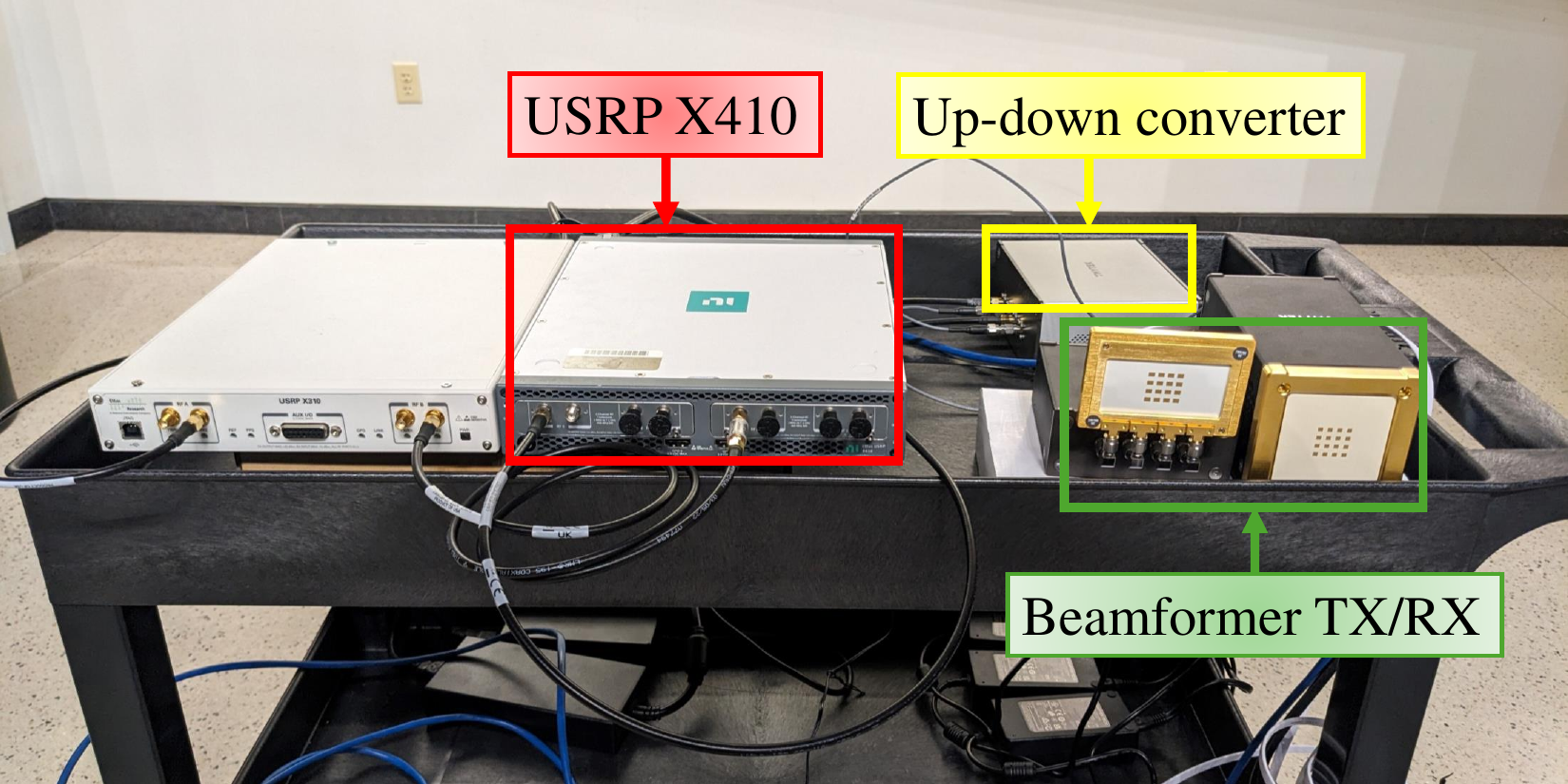}
       \label{fig:TX}
   }
   \subfigure[Communication RX.]
   {
      \includegraphics[width=0.45\columnwidth]{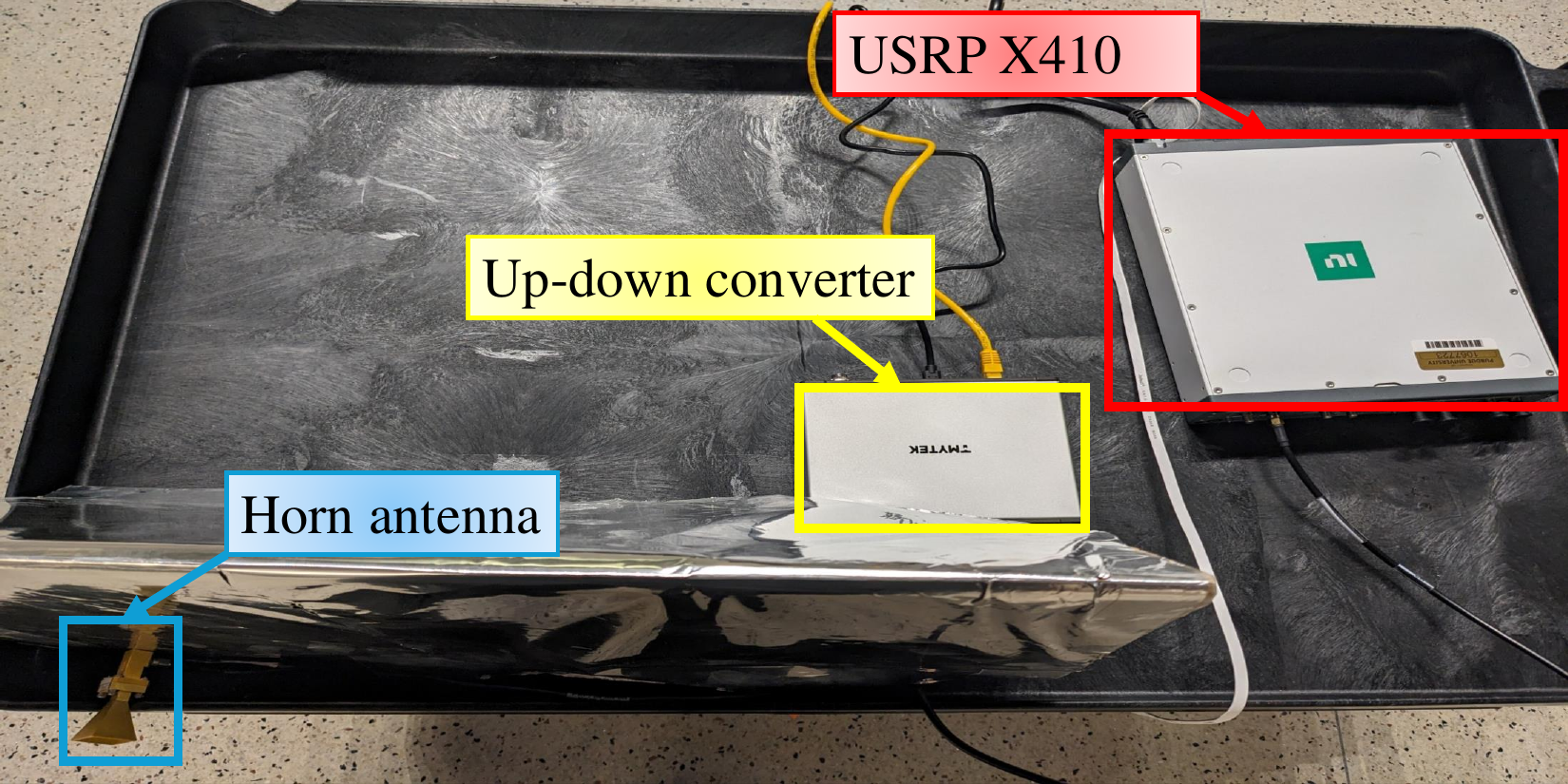}
      \label{fig:RX}
   }
   \subfigure[Radar estimation setup.]
   {
       \includegraphics[width=0.45\columnwidth]{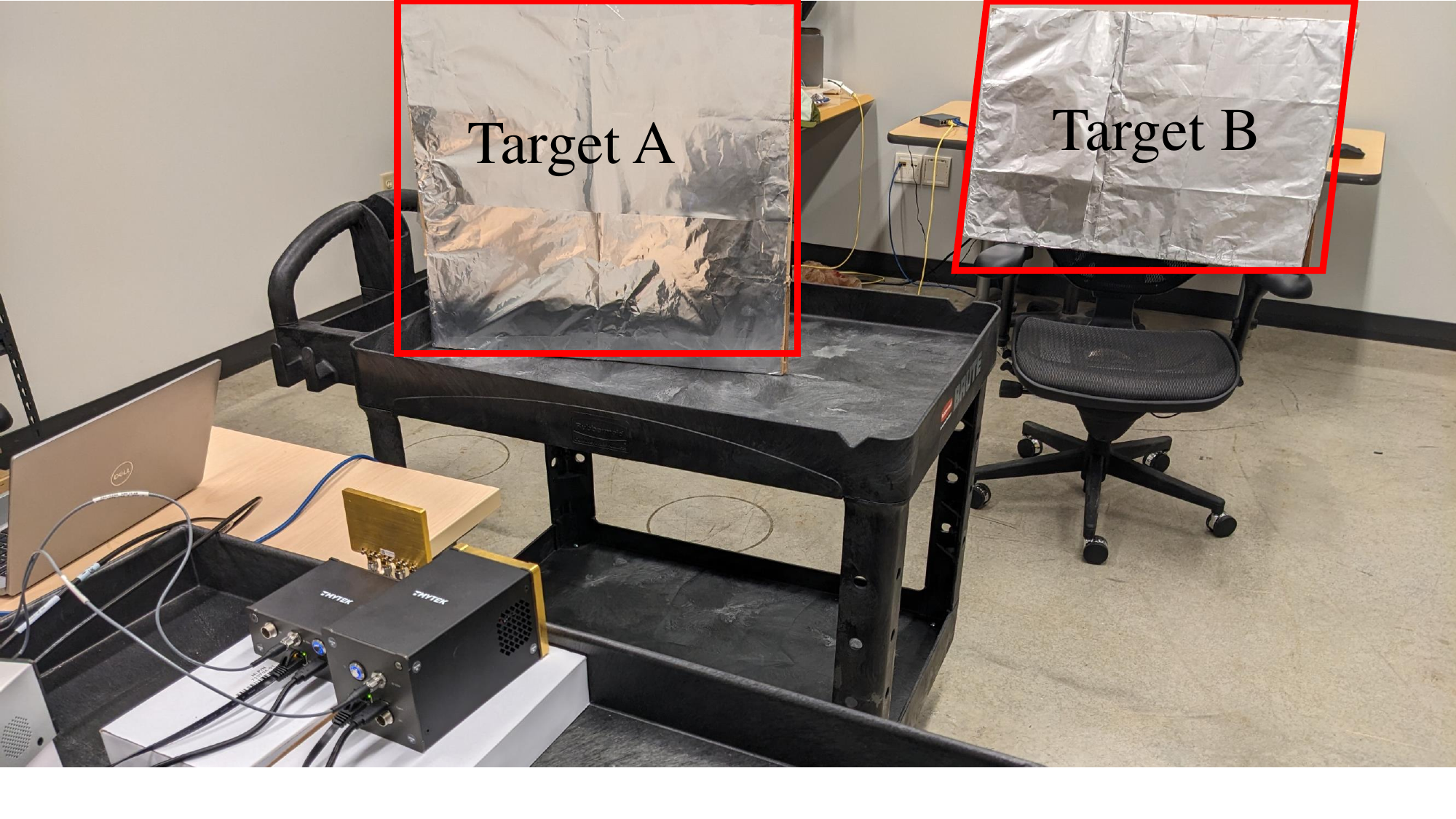}
       \label{fig:experiment_setting_radar}
   }
   \subfigure[Communication setup.]
   {
      \includegraphics[width=0.45\columnwidth]{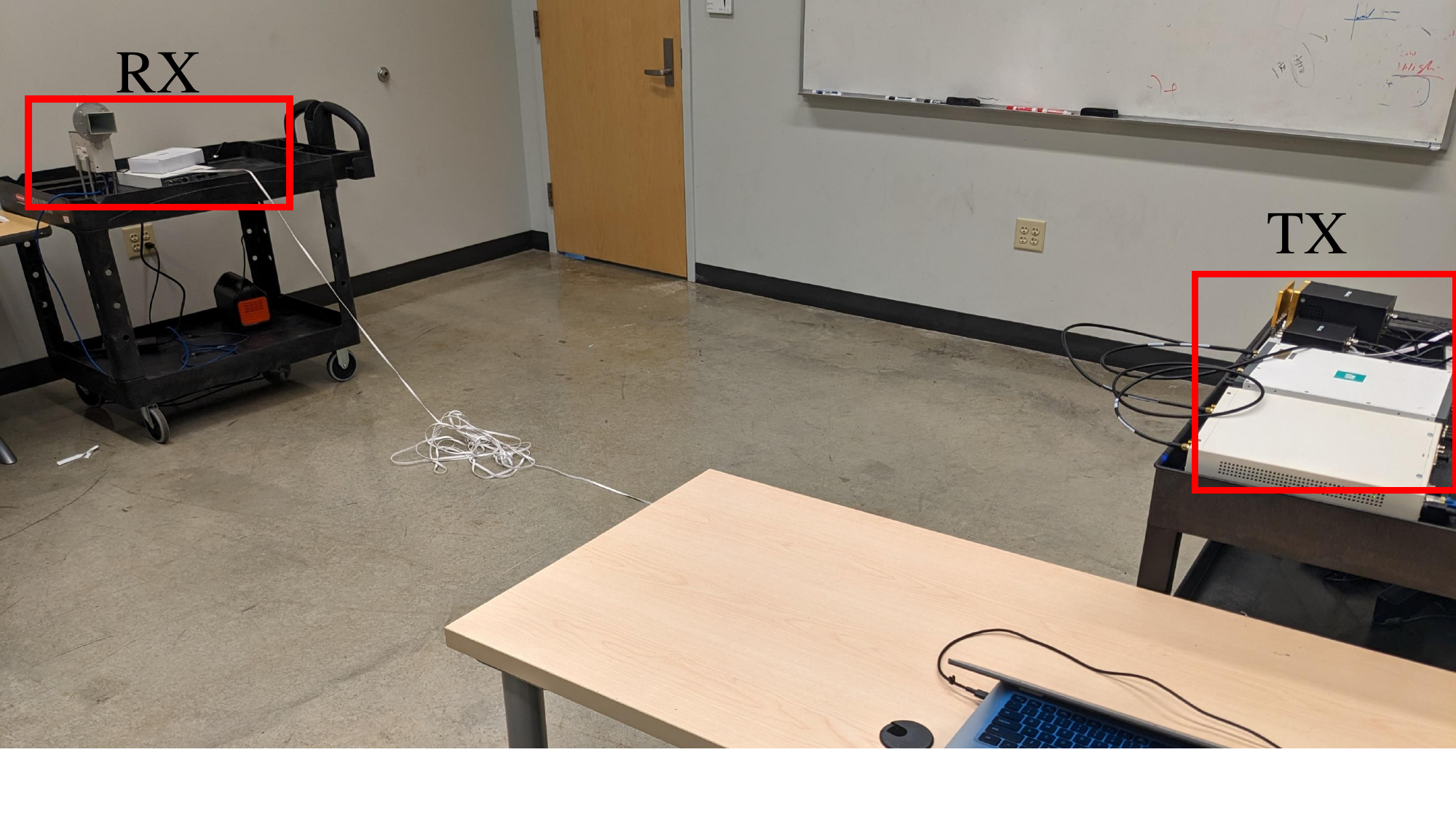}
      \label{fig:experiment_setting_comms}
   }
   \caption{Hardware and experimental setup.}
   \label{fig:experiment_setting}
\end{figure}

In this section, we further present experimental results to evaluate the performance of our system using an ISAC testbed.

\begin{figure}[!t]
 \centering
\includegraphics[width=2.6in, trim={0cm 0cm 0cm 0.2cm}, clip]{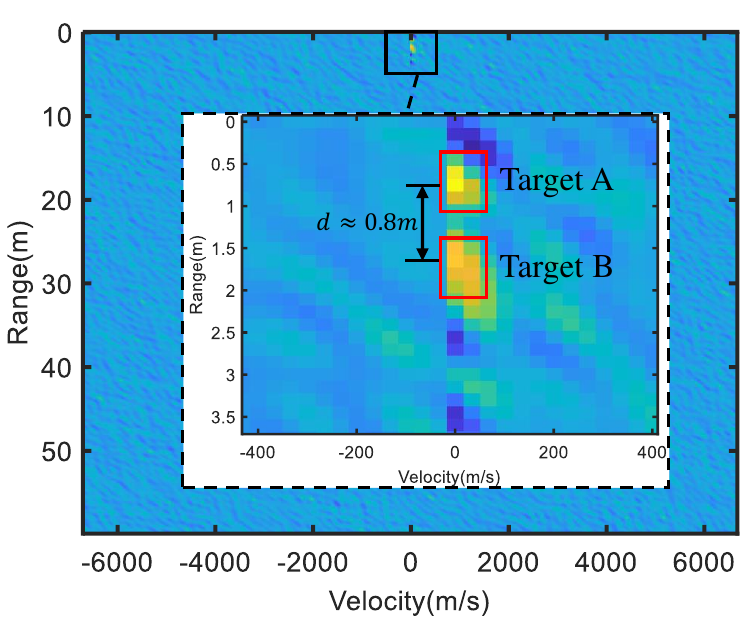}\\
 \caption{Experimental result of the range-velocity profile.}
 \label{fig:radar_experiment_result}
\end{figure}
 
\subsubsection{Hardware Implementation}

Our ISAC testbed, as shown in Figures~\ref{fig:TX} and \ref{fig:RX}, consists of two main components. The first component is the monostatic radar system, which also functions as the communication transmitter, as depicted in Fig.~\ref{fig:TX}. The second component is the communication receiver, as illustrated in Fig.~\ref{fig:RX}. For the operation of the system, two USRP X410 devices are deployed for generating and capturing the baseband OTFS signals. Additionally, up-down converters are integrated on both the transmitter and receiver sides to enable the conversion between the baseband and the 28~GHz radio frequency band. The radar system's front-end transceiver features dedicated beamformers for transmission (TX) and reception (RX). A horn antenna is deployed on the communication receiver side. The transmitter operates at the full sampling rate of 200~MHz, while the sampling rate for both the radar and communication receivers is reduced by a factor of 16.

\subsubsection{Radar Sensing Performance}

In the experiment, two RF reflectors are placed at a fixed distance of 0.8 m from each other, as shown in Fig.~\ref{fig:experiment_setting_radar}. Fig.~\ref{fig:radar_experiment_result} illustrates the range-velocity estimation under the previously described experimental conditions. Notably, even with the reduced sampling rate, the two closely located targets, A and B, can still be distinguished. This result demonstrates that our proposed method enables accurate target estimation with a reduced sampling rate, without compromising the sensing resolution.

\subsubsection{Communication performance}

\begin{figure}[!t]
 \centering
 \includegraphics[width=2.4in]{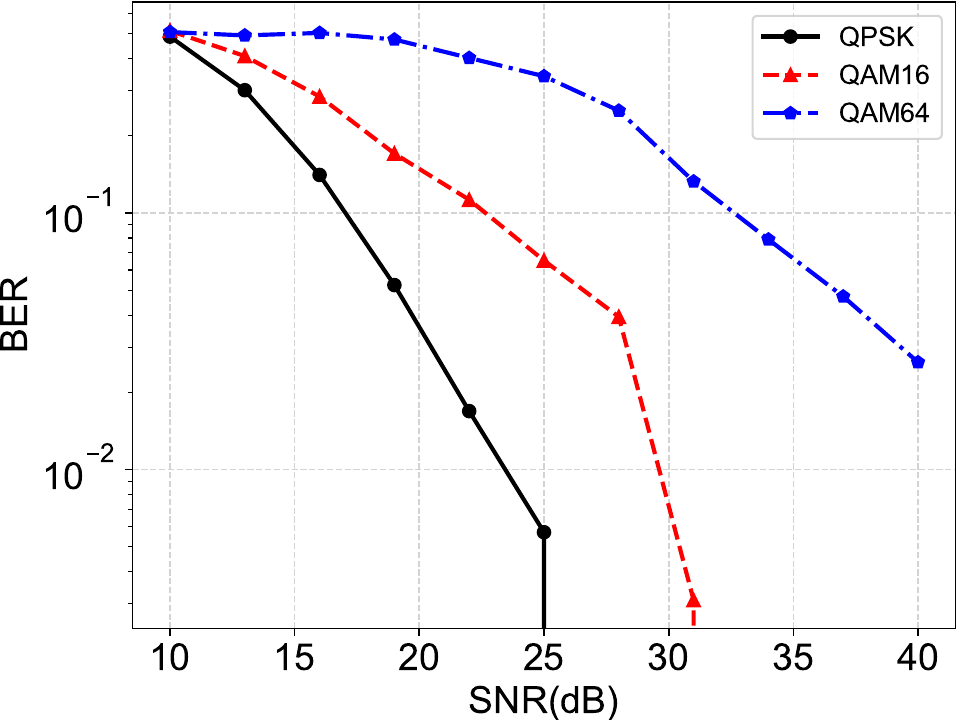}\\
 \caption{BER versus different modulation order with SDR testbed.}
 \label{fig:comms_experiment_result}
\end{figure}

We also evaluated the communication performance of our proposed system in a practical indoor environment. As illustrated in Fig.~\ref{fig:experiment_setting_comms}, the transmitter and receiver were positioned approximately 5 meters apart. By adjusting the transmission power at the transmitter side, we are able to manually control SNR to evaluate system performance under varying conditions.

The BER performance is depicted in Fig.~\ref{fig:comms_experiment_result}. The results demonstrate that the system performs well with QPSK modulation, achieving a BER of zero at an SNR of 25 dB under an uncoded scheme. For 16-QAM modulation, a BER of zero was observed at an SNR of approximately 30 dB. However, for 64-QAM modulation, the performance slightly degrades, likely due to hardware imperfections and potential signal distortion in the physical system. These practical experiments confirm the effectiveness of our system, highlighting its capability to maintain reliable communication performance under real-world conditions.

\section{Conclusion}\label{sec:conclusion}

This paper presents a novel OTFS-based ISAC system that
 addresses the challenges of high sampling rates and storage requirements in high-resolution radar estimation while simultaneously enabling effective communication. By strategically placing pilot symbols in the delay-Doppler domain and leveraging aliasing effects, our approach enables accurate radar estimation and efficient data transmission at sub-Nyquist sampling rates. A code-based spreading technique ensures the unambiguous sensing range, while an iterative interference cancellation algorithm enhances radar accuracy by mitigating data-induced interference. Furthermore, we developed a comprehensive transceiver pipeline optimized for reduced sampling rates, incorporating synchronization, iterative channel estimation, and data detection to ensure reliable communication. Experimental validation using an SDR-based ISAC testbed confirms that our system significantly lowers sampling requirements while maintaining sensing resolution and communication performance. This work provides a cost-effective solution for advancing integrated sensing and communication technologies, paving the way for practical deployment in next-generation wireless networks.

\bibliographystyle{IEEEtran}
\bibliography{main}

\end{document}